\documentclass[a4paper,11pt]{article}

\usepackage[top=1.0in,right=1.0in,left=1.0in,bottom=1.75in]{geometry}

\usepackage{latexsym}
\usepackage{amssymb}
\usepackage{float}
\usepackage{graphics,graphicx}
\usepackage{tabularx}
\usepackage{amsfonts}
\usepackage{amsmath}
\usepackage{enumerate}
\usepackage{booktabs}
\usepackage{multirow}
\usepackage{sectsty}
\usepackage[affil-it,auth-sc]{authblk}
\usepackage{subfig}

\sectionfont{\normalsize}
\subsectionfont{\normalsize}

\begin{document}

\newcommand{\singlespace}{\baselineskip=12pt\lineskiplimit=0pt\lineskip=0pt}
\def\ds{\displaystyle}

\newcommand{\beq}{\begin{equation}}
\newcommand{\eeq}{\end{equation}}
\newcommand{\lb}{\label}
\newcommand{\ph}{\phantom}
\newcommand{\beqar}{\begin{eqnarray}}
\newcommand{\eeqar}{\end{eqnarray}}
\newcommand{\barr}{\begin{array}}
\newcommand{\earr}{\end{array}}
\newcommand{\jump}{\parallel}
\newcommand{\Ehat}{\hat{E}}
\newcommand{\That}{\hat{\bf T}}
\newcommand{\Ahat}{\hat{A}}
\newcommand{\chat}{\hat{c}}
\newcommand{\shat}{\hat{s}}
\newcommand{\khat}{\hat{k}}
\newcommand{\muhat}{\hat{\mu}}
\newcommand{\mc}{M^{\scriptscriptstyle C}}
\newcommand{\mei}{M^{\scriptscriptstyle M,EI}}
\newcommand{\mec}{M^{\scriptscriptstyle M,EC}}
\newcommand{\hbeta}{{\hat{\beta}}}
\newcommand{\rec}[2]{\left( #1 #2 \ds{\frac{1}{#1}}\right)}
\newcommand{\rep}[2]{\left( {#1}^2 #2 \ds{\frac{1}{{#1}^2}}\right)}
\newcommand{\derp}[2]{\ds{\frac {\partial #1}{\partial #2}}}
\newcommand{\derpn}[3]{\ds{\frac {\partial^{#3}#1}{\partial #2^{#3}}}}
\newcommand{\dert}[2]{\ds{\frac {d #1}{d #2}}}
\newcommand{\dertn}[3]{\ds{\frac {d^{#3} #1}{d #2^{#3}}}}
\newcommand{\ct}{\captionof{table}}
\newcommand{\cf}{\captionof{figure}}

\def\c{{\circ}}
\def\bob{{\, \underline{\overline{\otimes}} \,}}
\def\ob{{\, \underline{\otimes} \,}}
\def\scalp{\mbox{\boldmath$\, \cdot \, $}}
\def\gdp{\makebox{\raisebox{-.215ex}{$\Box$}\hspace{-.778em}$\times$}}
\def\daa{\makebox{\raisebox{-.050ex}{$-$}\hspace{-.550em}$: ~$}}
\def\mK{\mbox{${\mathcal{K}}$}}
\def\cK{\mbox{${\mathbb {K}}$}}

\def\Xint#1{\mathchoice
   {\XXint\displaystyle\textstyle{#1}}%
   {\XXint\textstyle\scriptstyle{#1}}%
   {\XXint\scriptstyle\scriptscriptstyle{#1}}%
   {\XXint\scriptscriptstyle\scriptscriptstyle{#1}}%
   \!\int}
\def\XXint#1#2#3{{\setbox0=\hbox{$#1{#2#3}{\int}$}
     \vcenter{\hbox{$#2#3$}}\kern-.5\wd0}}
\def\ddashint{\Xint=}
\def\fpint{\Xint=}
\def\dashint{\Xint-}
\def\cpvint{\Xint-}
\def\intl{\int\limits}
\def\cpvintl{\cpvint\limits}
\def\fpintl{\fpint\limits}
\def\ointl{\oint\limits}
\def\bA{{\bf A}}
\def\ba{{\bf a}}
\def\bB{{\bf B}}
\def\bb{{\bf b}}
\def\bc{{\bf c}}
\def\bC{{\bf C}}
\def\bD{{\bf D}}
\def\bE{{\bf E}}
\def\be{{\bf e}}
\def\bbf{{\bf f}}
\def\bF{{\bf F}}
\def\bG{{\bf G}}
\def\bg{{\bf g}}
\def\bi{{\bf i}}
\def\bH{{\bf H}}
\def\bK{{\bf K}}
\def\bL{{\bf L}}
\def\bM{{\bf M}}
\def\bN{{\bf N}}
\def\bn{{\bf n}}
\def\bm{{\bf m}}
\def\b0{{\bf 0}}
\def\bo{{\bf o}}
\def\bX{{\bf X}}
\def\bx{{\bf x}}
\def\bP{{\bf P}}
\def\bp{{\bf p}}
\def\bQ{{\bf Q}}
\def\bq{{\bf q}}
\def\bR{{\bf R}}
\def\bS{{\bf S}}
\def\bs{{\bf s}}
\def\bT{{\bf T}}
\def\bt{{\bf t}}
\def\bU{{\bf U}}
\def\bu{{\bf u}}
\def\bv{{\bf v}}
\def\bw{{\bf w}}
\def\bW{{\bf W}}
\def\by{{\bf y}}
\def\bz{{\bf z}}
\def\T{{\bf T}}
\def\Te{\textrm{T}}
\def\Id{{\bf I}}
\def\bxi{\mbox{\boldmath${\xi}$}}
\def\balpha{\mbox{\boldmath${\alpha}$}}
\def\bbeta{\mbox{\boldmath${\beta}$}}
\def\bepsilon{\mbox{\boldmath${\epsilon}$}}
\def\bvarepsilon{\mbox{\boldmath${\varepsilon}$}}
\def\bomega{\mbox{\boldmath${\omega}$}}
\def\bphi{\mbox{\boldmath${\phi}$}}
\def\bsigma{\mbox{\boldmath${\sigma}$}}
\def\bfeta{\mbox{\boldmath${\eta}$}}
\def\bDelta{\mbox{\boldmath${\Delta}$}}
\def\btau{\mbox{\boldmath $\tau$}}
\def\tr{{\rm tr}}
\def\dev{{\rm dev}}
\def\div{{\rm div}}
\def\Div{{\rm Div}}
\def\Grad{{\rm Grad}}
\def\grad{{\rm grad}}
\def\Lin{{\rm Lin}}
\def\Sym{{\rm Sym}}
\def\Skw{{\rm Skew}}
\def\abs{{\rm abs}}
\def\Re{{\rm Re}}
\def\Im{{\rm Im}}
\def\capB{\mbox{\boldmath${\mathsf B}$}}
\def\capC{\mbox{\boldmath${\mathsf C}$}}
\def\capD{\mbox{\boldmath${\mathsf D}$}}
\def\capE{\mbox{\boldmath${\mathsf E}$}}
\def\capG{\mbox{\boldmath${\mathsf G}$}}
\def\tcapG{\tilde{\capG}}
\def\capH{\mbox{\boldmath${\mathsf H}$}}
\def\capK{\mbox{\boldmath${\mathsf K}$}}
\def\capL{\mbox{\boldmath${\mathsf L}$}}
\def\capM{\mbox{\boldmath${\mathsf M}$}}
\def\capR{\mbox{\boldmath${\mathsf R}$}}
\def\capW{\mbox{\boldmath${\mathsf W}$}}

\def\i{\mbox{${\mathrm i}$}}
\def\mC{\mbox{\boldmath${\mathcal C}$}}
\def\mB{\mbox{${\mathcal B}$}}
\def\mE{\mbox{${\mathcal{E}}$}}
\def\mL{\mbox{${\mathcal{L}}$}}
\def\mK{\mbox{${\mathcal{K}}$}}
\def\mV{\mbox{${\mathcal{V}}$}}
\def\C{\mbox{\boldmath${\mathcal C}$}}
\def\E{\mbox{\boldmath${\mathcal E}$}}

\def\AAM{{\it Advances in Applied Mechanics }}
\def\ACME{{\it Arch. Comput. Meth. Engng.}}
\def\ARMA{{\it Arch. Rat. Mech. Analysis}}
\def\AMR{{\it Appl. Mech. Rev.}}
\def\ASCEEM{{\it ASCE J. Eng. Mech.}}
\def\ACTA{{\it Acta Mater.}}
\def\CMAME {{\it Comput. Meth. Appl. Mech. Engrg.}}
\def\CRAS{{\it C. R. Acad. Sci. Paris}}
\def\CRM{{\it Comptes Rendus M\'ecanique}}
\def\EFM{{\it Eng. Fracture Mechanics}}
\def\EJMA{{\it Eur.~J.~Mechanics-A/Solids}}
\def\IJES{{\it Int. J. Eng. Sci.}}
\def\IJF{{\it Int. J. Fracture}}
\def\IJMS{{\it Int. J. Mech. Sci.}}
\def\IJNAMG{{\it Int. J. Numer. Anal. Meth. Geomech.}}
\def\IJP{{\it Int. J. Plasticity}}
\def\IJSS{{\it Int. J. Solids Structures}}
\def\IngA{{\it Ing. Archiv}}
\def\JAM{{\it J. Appl. Mech.}}
\def\JAP{{\it J. Appl. Phys.}}
\def\JAE{{\it J. Aerospace Eng.}}
\def\JE{{\it J. Elasticity}}
\def\JM{{\it J. de M\'ecanique}}
\def\JMPS{{\it J. Mech. Phys. Solids}}
\def\JSV{{\it J. Sound and Vibration}}
\def\MACRO{{\it Macromolecules}}
\def\MMT{{\it Mech. Mach. Th.}}
\def\MOM{{\it Mech. Materials}}
\def\MMS{{\it Math. Mech. Solids}}
\def\MMT{{\it Metall. Mater. Trans. A}}
\def\MPCPS{{\it Math. Proc. Camb. Phil. Soc.}}
\def\MSE{{\it Mater. Sci. Eng.}}
\def\NATURE{{\it Nature}}
\def\NATUREM{{\it Nature Mater.}}
\def\PHIL{{\it Phil. Trans. R. Soc.}}
\def\PMPS{{\it Proc. Math. Phys. Soc.}}
\def\PNAS{{\it Proc. Nat. Acad. Sci.}}
\def\PRE{{\it Phys. Rev. E}}
\def\PRL{{\it Phys. Rev. Letters}}
\def\PRSL{{\it Proc. R. Soc.}}
\def\RIIT{{\it Rozprawy Inzynierskie - Engineering Transactions}}
\def\ROCK{{\it Rock Mech. and Rock Eng.}}
\def\QAM{{\it Quart. Appl. Math.}}
\def\QJMAM{{\it Quart. J. Mech. Appl. Math.}}
\def\SCIENCE{{\it Science}}
\def\SCRMAT{{\it Scripta Mater.}}
\def\SM{{\it Scripta Metall.}}
\def\ZAMM{{\it Z. Angew. Math. Mech.}}
\def\ZAMP{{\it Z. Angew. Math. Phys.}}
\def\ZVDI{{\it Z. Verein. Deut. Ing.}}

\renewcommand\Affilfont{\itshape}
\setlength{\affilsep}{1em}
\renewcommand\Authsep{, }
\renewcommand\Authand{ and }
\renewcommand\Authands{ and }
\setcounter{Maxaffil}{2}

\title{
Effects of constraint curvature on structural instability: \\
tensile buckling and multiple bifurcations
}
\author[1]{D. Bigoni\footnote{Corresponding author:\,e-mail:\,bigoni@ing.unitn.it; phone:\,+39\,0461\,282507.}}
\author[1]{D. Misseroni}
\author[1]{G. Noselli}
\author[2]{D. Zaccaria}
\affil[1]{
Department of Mechanical and Structural Engineering, University of Trento
via~Mesiano~77, Trento, Italy.
}
\affil[2]{
Department of Civil and Environmental Engineering, University of Trieste
piazzale~Europa~1, Trieste, Italy.
}

\date{}
\maketitle

\begin{abstract}

\noindent Bifurcation of an elastic structure crucially depends on the curvature of the constraints against which the ends of the structure are prescribed to move, an
effect which deserves more attention than it has received so far. In fact,
we show theoretically and we provide definitive experimental verification that an appropriate curvature of the constraint over which the end of a structure has to slide strongly affects buckling loads and can induce: (i.) tensile buckling;
(ii.) decreasing- (softening), increasing- (hardening), or constant-load (null stiffness) postcritical behaviour; (iii.) multiple bifurcations, determining for instance two
bifurcation loads (one tensile and one compressive) in a single-degree-of-freedom elastic system.
We show how to design a constraint profile to obtain a desired postcritical behaviour and
we provide the solution for the elastica constrained to slide along a circle on one end, representing the first example of an inflexional elastica developed from a buckling in tension.
These results have important practical implications
in the design of compliant mechanisms and may find applications in devices operating in quasi-static or dynamic conditions.

\end{abstract}


\section{Introduction}\lb{INTRO}

\begin{figure}[!htcb]
\renewcommand{\figurename}{\footnotesize{Fig.}}
    \begin{center}
    \includegraphics[width = 9 cm]{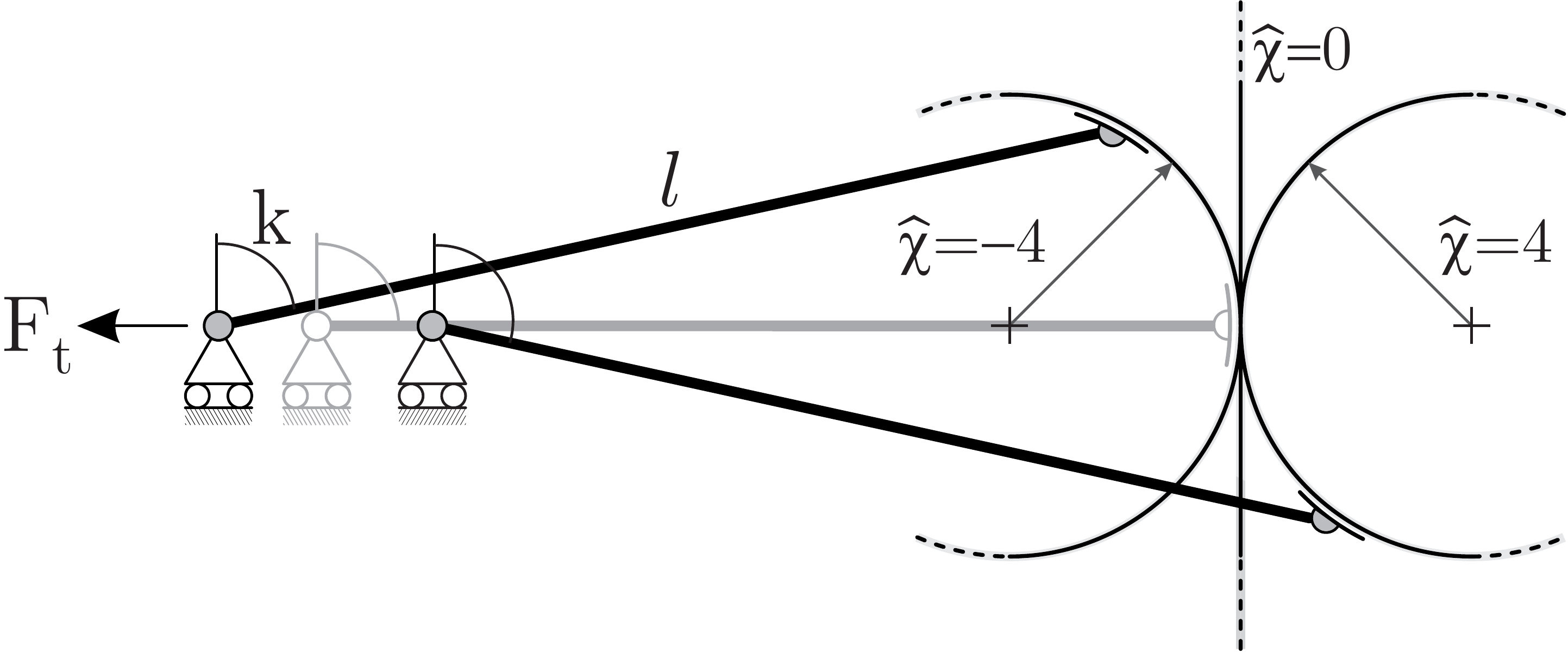}
    \caption{\footnotesize A one-degree-of-freedom structure (with a rotational elastic spring at its left end) evidencing compressive or tensile buckling as a function of the curvature of the constraint (a circular profile with constant curvature, $\widehat{\chi}= \pm 4$)
on which the hinge on the right of the structure has to slide.}
    \label{1DOF}
    \end{center}
\end{figure}

We begin with a simple example, by considering a one-degree-of-freedom elastic structure made up of a rigid rod connected with a rotational linear elastic spring on its
left end and with a pin constrained to move on a circle (of radius $R_c$)
centered on the rod's axis on the right (Fig.~\ref{1DOF}).
The structure is subject to a horizontal force, so that when this load is compressive and the circle degenerates to a line (null curvature),
the structure buckles at the compressive force $F = -k/l$.
Our interest is to analyze the case when the curvature of the constraint is not null, revealing
that this curvature strongly affects the critical load,
which results to be a {\it tensile} force\footnote{
Tensile buckling of an elastic structure governed by the elastica (in which all elements are subject to tension) has been recently discovered by Zaccaria {\it et al.} (2011).
}
in the negative curvature case ($F_t=k/3/l$, for $\widehat{\chi}=l/R_c= -4$) and a compressive load
for positive curvature ($F_c=-k/5/l$, for $\widehat{\chi}=l/R_c=4$).

The example shows that the curvature of the constraint at the end of a structure deeply affects its critical loads\footnote{
The fact that the curvature influences the critical load was observed in different terms already by Timoshenko and Gere (1936), who analyzed the case of the so-called \lq load through a fixed point'.
However, they did not generalize the problem enough to discover that: tensile buckling, multiple bifurcations and inflexional tensile elastica during the postcritical behaviour can be obtained, which  is the topic
attacked in the present article.
},
but also
the shape of the curve defining the constraint influences the postcritical behaviour, which displays a rising-load (hardening) behaviour in the case of null curvature and a
decreasing-load (softening) behaviour for circular profiles (for instance when
$\widehat{\chi} = \pm 4$ as in the structure shown in Fig. \ref{1DOF}). Moreover, the postcritical behaviour connected to the tensile (compressive) bifurcation is rather peculiar, since the tensile
(compressive) force
needed to buckle the structure decreases until it vanishes and becomes compressive (tensile), during continued displacement of the structure end.

Once the lesson on the curvature and the shape of the constraint is clear, it becomes easy to play with these structural elements and discover several new effects. Some of these are listed in the following.

\begin{itemize}

\item A constraint profile can be designed to provide a \lq hardening', \lq softening' or even a \lq {\it neutral}' postcritical behaviour, where the displacement grows at constant load.
More in general, a constraint profile can be designed to obtain a variety of desired postcritical behaviours, including situations in which the stability of the path changes during postcritical deformation.

\item A negative and a positive curvature can be combined in an \lq S-shaped constraint'
(see the inset of Fig.~\ref{post_1DOF}) to yield {\it  a
one-degree-of-freedom structure with two buckling loads, one tensile and one compressive}.
\begin{figure}[!htcb]
\renewcommand{\figurename}{\footnotesize{Fig.}}
    \begin{center}
    \includegraphics[width = 14 cm]{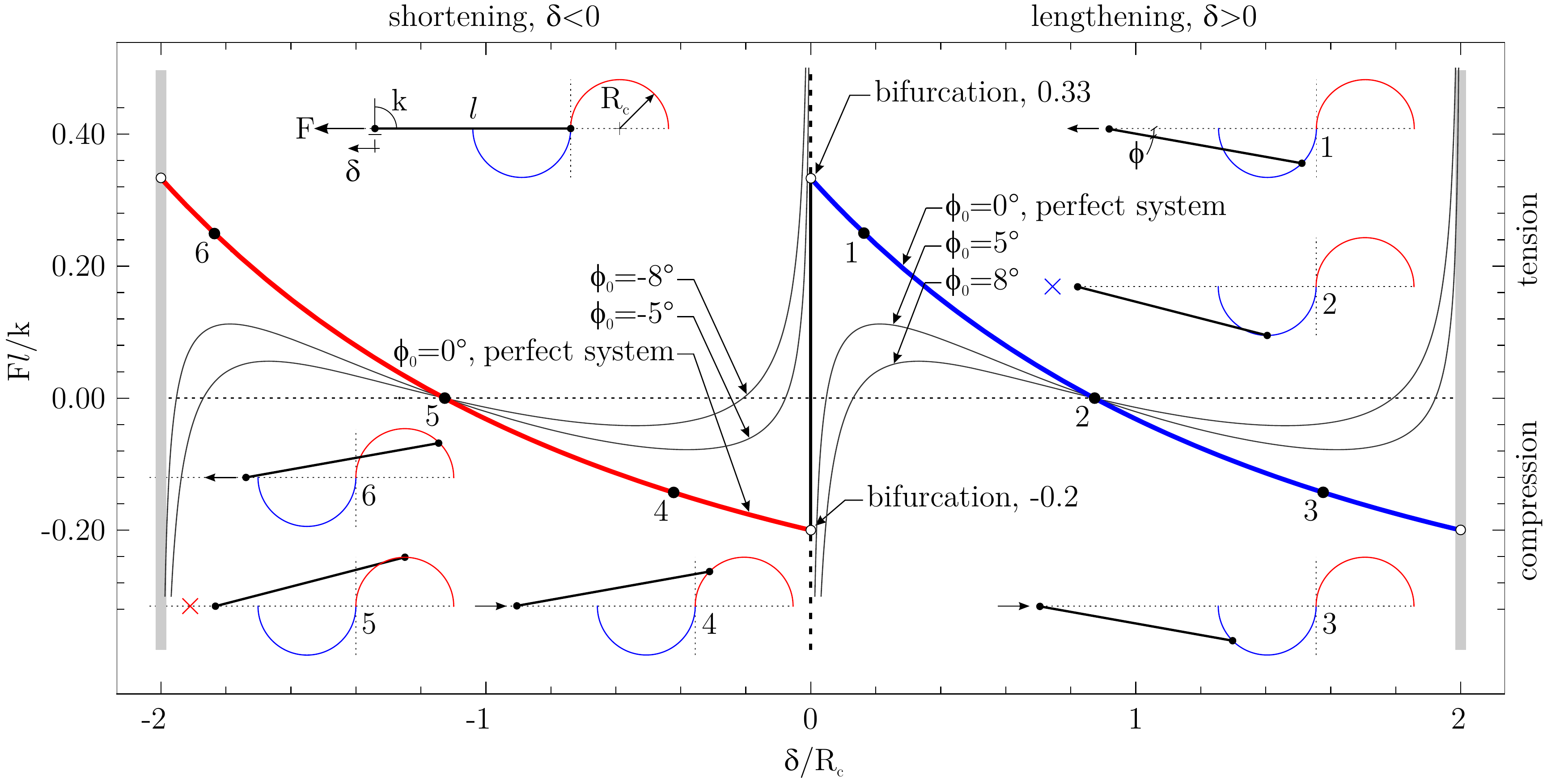}
    \caption{\footnotesize The behaviour of a one-degree-of-freedom structure evidencing two buckling loads, one compressive and one tensile. These are the effect of the discontinuity in the curvature
of the (piecewise circular) constraint. Note that: (i.) the two unstable postcritical branches are identical modulo a horizontal shift and that (ii.) during the postcritical behaviour there is a transition from tensile (compressive) to compressive (tensile) load.}
    \label{post_1DOF}
    \end{center}
\end{figure}

\end{itemize}

In the case of the \lq S-shaped constraint', imperfections suppress bifurcations and the stability of the equlibrium path strongly depends on the {\it sign} of the imperfection. For tensile forces,
if the imperfection has a positive sign ($\phi_0>0$), the
equilibrium path of the system becomes unstable after a peak in the load is reached, while if the sign is negative ($\phi_0<0$), the structure remains in a metastable equilibrium
configuration which asymptotically approaches an unstable configuration (Fig.~\ref{post_1DOF}).

Finally, we can now appreciate the role played by the curvature of a constraint in the more interesting case of a structural element governed by the elastica, a research aspect passed unnoticed until now.
In this article, we show that consideration of this curvature provides a generalization of the findings by Zaccaria {\it et al.} (2011), so that their \lq slider' can be seen as a special case
of the curved constraint introduced in the present article and the elastica developing after a tensile buckling is of inflexional type, while that investigated by Zaccaria {\it et al.} (2011) is
non-inflexional.
We fully develop the theory of the elastica constrained to slide along a circle on one of its ends and we experimentally confirm the theoretical findings with experiments designed and realized by us
at the Laboratory for Physical Modeling of Structures and Photoelasticity.

The article is organized as follows. We begin presenting a generalization of the one-degree-of-freedom structure shown in Fig.~\ref{1DOF}, to highlight: (i.) the effects of the curvature of the constraint,
(ii.) the multiplicity of bifurcation loads, (iii.) the behaviour of the imperfect system, and (iv.) the possibility of designing a constraint profile to obtain a given postcritical behaviour.
Later we introduce a continuos system, made up of an inextensible beam governed by the Euler elastica and we solve the critical
loads and the nonlinear postcritical large-deformation behaviour, through explicit integration of the elastica.
Throughout the text, we complement theoretical results with experiments confirming all our findings for discrete and continuous elastic systems.
A movie providing a simple illustration of the concepts exposed in this article, together with
a view of experimental results, is provided in the electronic supplementary material, see also
http://www.ing.unitn.it/dims/ssmg.php.

\section{Effect of constraint's curvature on a one-degree-of-freedom elastic structure}\lb{ODOF}

\begin{figure}[!htcb]
\renewcommand{\figurename}{\footnotesize{Fig.}}
    \begin{center}
    \includegraphics[width = 8 cm]{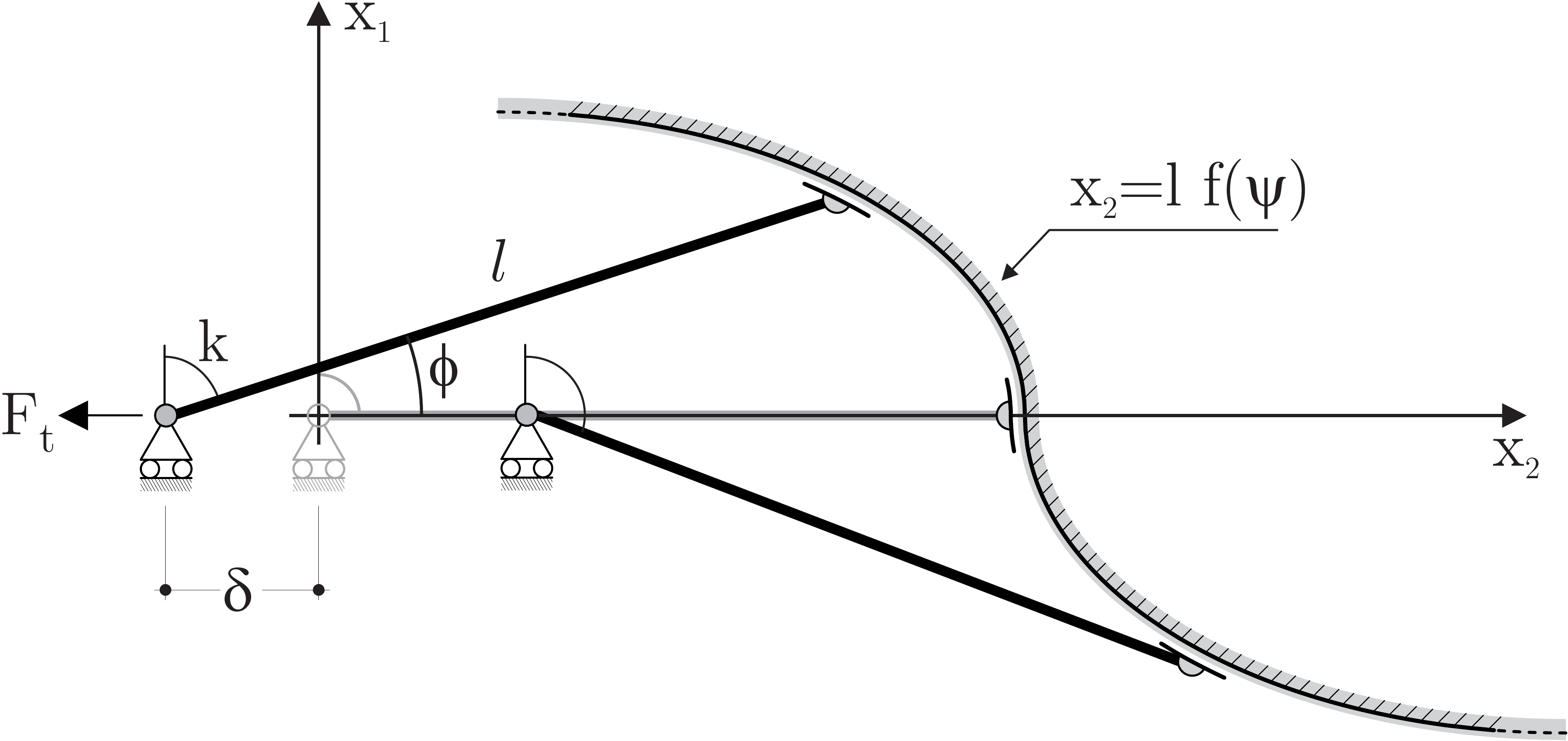}
    \caption{\footnotesize A one-degree-of-freedom structure with an hinge constrained to slide along a generic smooth profile at the right end and a rotational linear-elastic spring at the left end.}
    \label{1DOF_S}
    \end{center}
\end{figure}

Bifurcation load and equilibrium paths of the one-degree-of-freedom structure illustrated in Fig.~\ref{1DOF_S}
(where the constraint is assumed smooth and described in the $x_1$--$x_2$ reference system as $x_2 = l\,f(\psi)$, with $\psi=x_1/l \in[0,1]$ and $f'(0)=0$) can be calculated by considering a deformed mode
defined by the rotation angle $\phi$. Assuming a possible imperfection in terms of an initial inclination $\phi_0$, the elongation of the system and the potential energy are respectively
\beq
\delta = l\left[\cos \phi-\cos\phi_0 -f(\sin\phi) + f(\sin\phi_0)\right]
\eeq
and
\beq
W(\phi) = \frac{1}{2} k (\phi-\phi_0)^2 - F l \left[\cos \phi-\cos\phi_0 -f(\sin\phi) + f(\sin\phi_0) \right],
\eeq
so that solutions of the equilibrium problem are governed by
\beq
\lb{oreste_bursi}
F = -\frac{k\,(\phi-\phi_0)} {l[\sin \phi + \cos \phi \,f'(\sin\phi)]},
\eeq
where $f'= \partial f/\partial \psi$, so that
the critical load for the perfect system, $\phi_0=0$, is
\beq
F_{cr} =- \frac{k}{l[1+ f''(0)]},
\eeq
where, since $f'(0)=0$, $f''(0)=\widehat{\chi}(0)$ is the signed curvature at $\phi=0$.
Stability can be judged on the basis of the sign of the second derivative of the potential energy
\beq
\frac{\partial^2 W(\phi)}{\partial \phi^2} = k + Fl \left(\cos \phi - f' \sin\phi  + f'' \cos^2\phi  \right),
\eeq
showing that the trivial configuration of the perfect system is always unstable beyond the critical load.

In the case when the profile of the constraint is a circle\footnote{Note that in the case of a circle the dimensionless signed curvature is $\widehat{\chi} = \pm l/R_c$, with $l$
being the length of the rigid bar and $R_c$ the radius of the circle.} of dimensionless radius $1/|\widehat{\chi}|=1/l|\chi|$  as in Fig.~\ref{1DOF}, the non-trivial equilibrium configurations are
given by
\beq
\lb{cerchio_nel_grano}
F = - \frac{k\,(\phi-\phi_0)\sqrt{1 - \widehat{\chi}^2 \sin^2\phi}} {l\,\sin \phi (\widehat{\chi} \cos \phi + \sqrt{1 - \widehat{\chi}^2 \sin^2\phi})},
\eeq
and result to be
stable when
\beq
\lb{cerchio_nel_fieno}
1 - \widehat{\chi}^2 \sin^2\phi -(\phi-\phi_0)( \cot \phi - \widehat{\chi} \sin \phi \sqrt{1- \widehat{\chi}^2 \sin^2\phi})>0 .
\eeq
Eqs.~(\ref{cerchio_nel_grano}) and (\ref{cerchio_nel_fieno}) have been used to solve the special case of Fig.~\ref{1DOF} ($\widehat{\chi} = \pm4$), with an \lq S-shaped' constraint (so that $\widehat{\chi}$ is discontinuous at $\phi=0$), to obtain the results plotted in Fig.~\ref{post_1DOF}.

\subsection{The design of the postcritical behaviour}

It is important to emphasize that {\it the shape of the profile on which one end of the structure has to slide can be designed to obtain \lq desired postcritical behaviours'}.
Let us assume that we want to obtain a certain force-displacement $F/\delta$ postcritical behaviour. Since
\beq
\delta = l\left[\sqrt{1-\psi^2} -f(\psi) \right],
\eeq
to assume a certain $F/\delta$ relation is equivalent to assume a given dependence of $F$ on $\psi$; therefore we introduce the dimensionless function
\beq
\beta(\psi) = \frac{l}{k}\, F(\delta(\psi)).
\eeq
Employing Eq.~(\ref{oreste_bursi}) we obtain the condition
\beq
f(\psi) = \sqrt{1-\psi^2} - \int_0^\psi \frac{\arcsin\gamma}{\beta(\gamma)\,\sqrt{1-\gamma^2}} \,d\gamma \,,
\eeq
satisfying $f(0) = 1$ and $f'(0)=0$.

\begin{figure}[H]
\renewcommand{\figurename}{\footnotesize{Fig.}}
    \begin{center}
    \includegraphics[width = 12 cm]{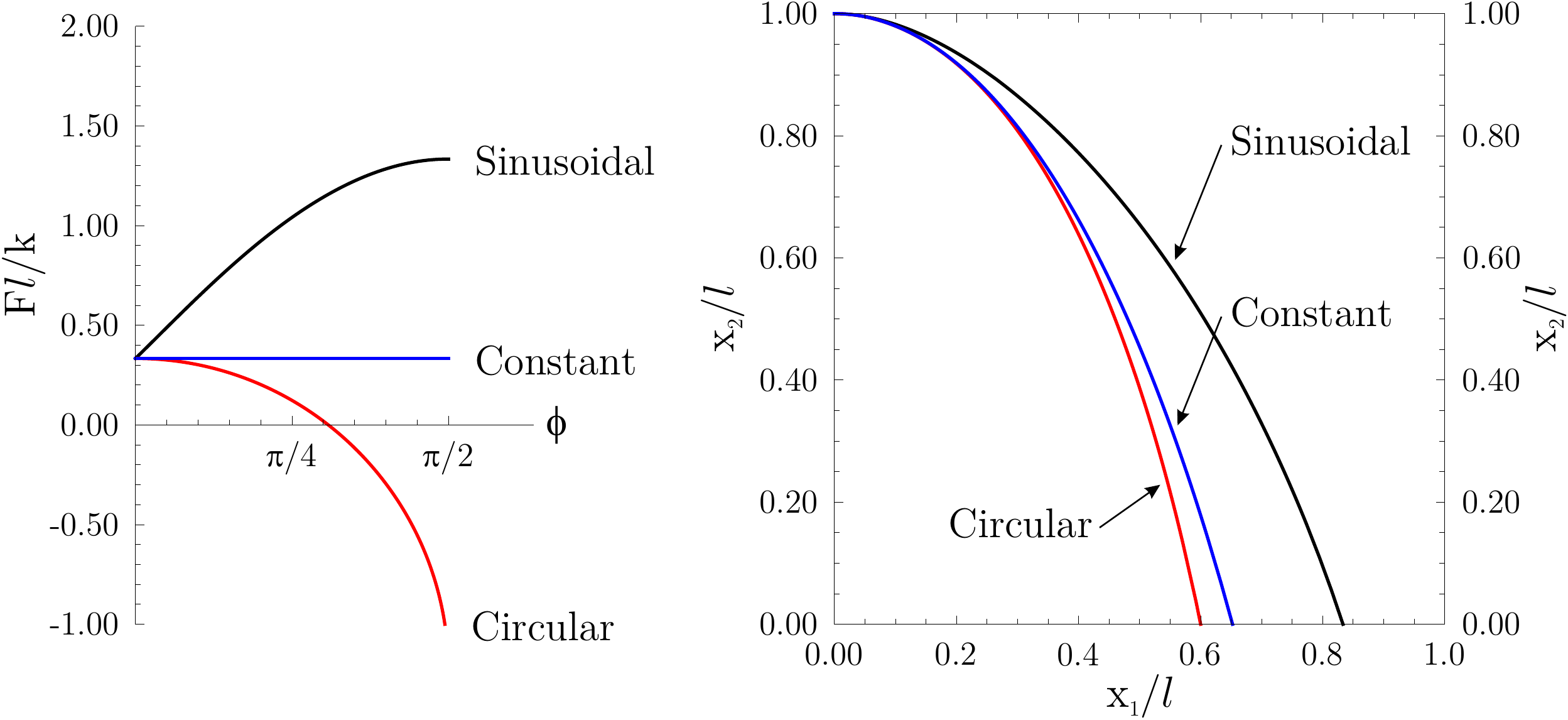}
    \caption{\footnotesize Designed profiles (on the right) to obtain a given
force-rotation postcritical response (on the left). The postcritical responses, given in terms of dimensionless force versus rotation of the structure are:
sinusoidal, circular and constant (or \lq neutral').}
    \label{profili}
    \end{center}
\end{figure}

Some profile designed to obtain particular force $F$ versus rotation $\phi$ postcritical behaviours (a sinusoidal, a circular and a constant) are sketched in Fig.~\ref{profili}.
An interesting case is that of the neutral (or constant) postcritical behaviour, in which the rotation $\phi$ (and therefore also the displacement) can
grow at constant load\footnote{
A neutral postcritical behaviour has been found also by G\'{a}sp\'{a}r (1984) employing a structural model completely different from that considered by us.
},
which can be obtained employing the constraint profile expressed as
\beq
f(\psi)=\sqrt{1-\psi^2}- \frac{1}{2 \beta} \Big(\arcsin \psi\Big)^2~~\mbox{where}~~ \beta=\frac{F_{cr} l}{k}.
\eeq

\subsection{Experiments on one-degree-of-freedom elastic systems: multiple buckling and neutral postcritical response}\lb{exp-123}

\begin{figure}[!htcb]
\renewcommand{\figurename}{\footnotesize{Fig.}}
    \begin{center}
    \includegraphics[width = 12 cm]{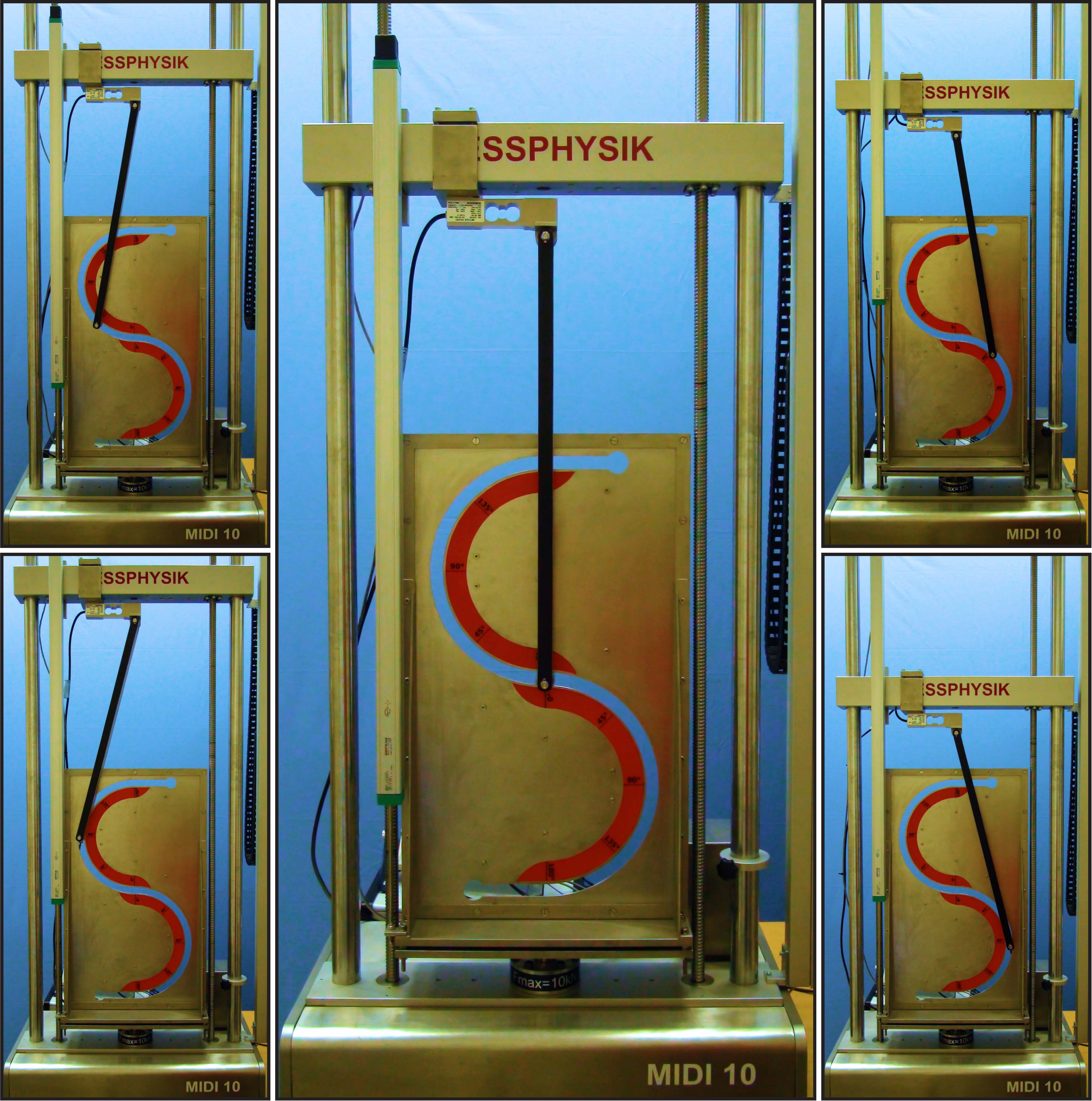}
    \caption{\footnotesize Experimental set-up for the \lq S-shaped' structure with a groove corresponding to two circles.
Two photos taken during elongation (shortening) are reported on the left (on the right).}
    \label{ungrado}
    \end{center}
\end{figure}

\begin{figure}[!htcb]
\renewcommand{\figurename}{\footnotesize{Fig.}}
    \begin{center}
    \includegraphics[width = 12 cm]{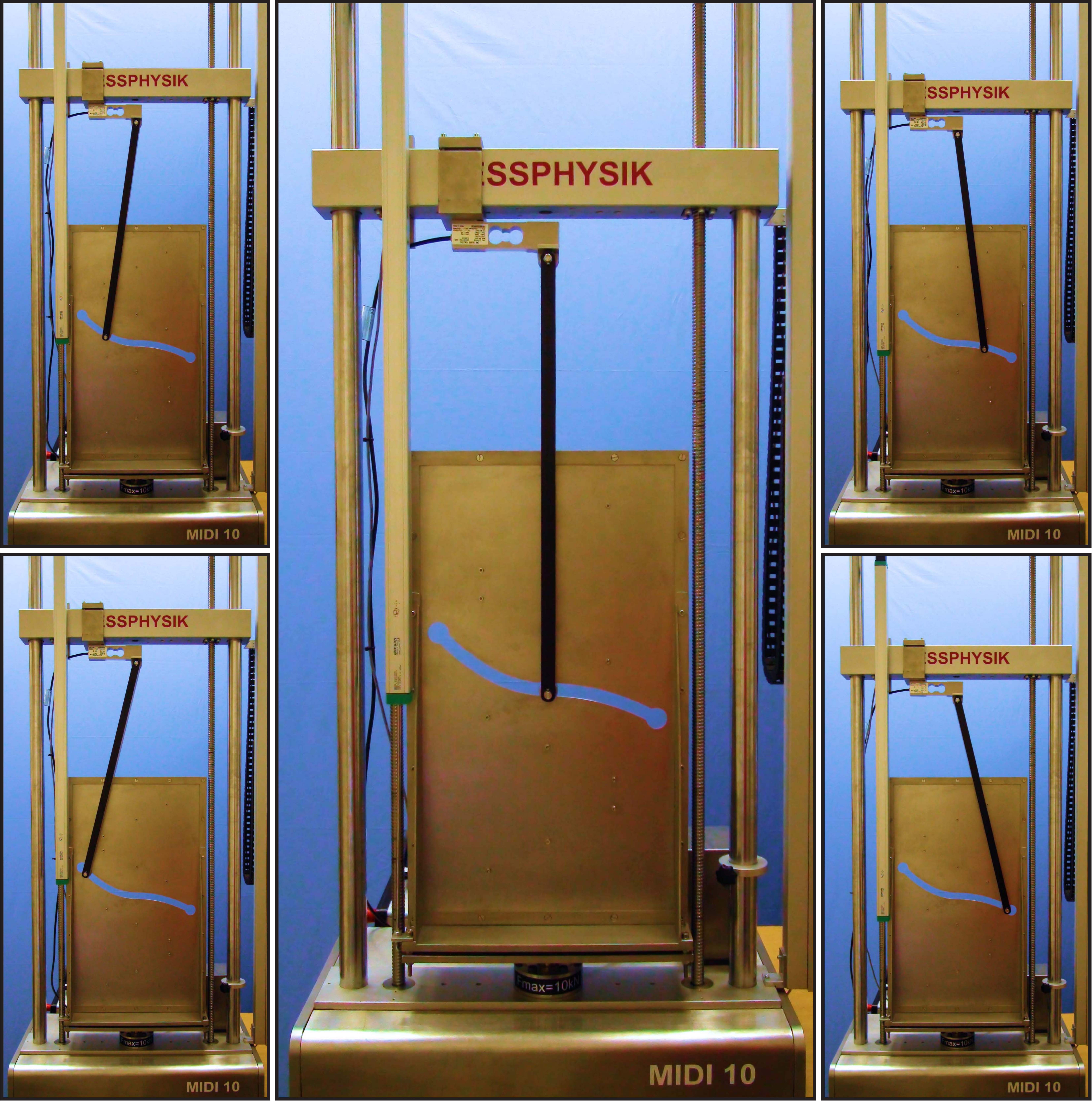}
    \caption{\footnotesize Experimental set-up for the structure providing the neutral postcritical response. Two photos taken during elongation (shortening) are reported on the left (on the right).}
    \label{ungrado2}
    \end{center}
\end{figure}

The behaviours obtained employing the simple one-degree-of-freedom structures are not a mathematical curiosity, but can be realized in practice.
In particular, we have realized the \lq S-shaped' circular constraint shown in Fig.~\ref{1DOF} and the profile illustrated in Fig.~\ref{profili} (on the right, labelled \lq constant'), the latter to show a \lq neutral' or, in other words \lq constant-force',   response.
The experimental apparatuses are shown in Fig.~\ref{ungrado} and in Fig.~\ref{ungrado2} (the former relative to semi-circular profiles, the latter to the profile providing the neutral post-critical response),
where the
grooves have been laser cut (by HTR Laser \& Water cut, BZ, Italy) in a 2 mm thick plate of AISI 304 steel and the roller has been realized with a (17 mm diameter) steel cylinder mounted with two roller bearings (SKF-61801-2Z).
The rigid bars 600 mm $\times$ 25 mm $\times$ 20 mm have been machined from an aluminum bar and lightened with
longitudinal groowes (see Appendix A), so that the final weight is 820\,gr.
The elastic hinge has been realized with three identical rotational springs have been employed, which have been designed using equations (32) of Brown (1981) and realized in (4 mm diameter) music wire ASTM A228,
see Appendix A for further details on experiments.

\begin{figure}[!htcb]
\renewcommand{\figurename}{\footnotesize{Fig.}}
    \begin{center}
    \includegraphics[width = 12 cm]{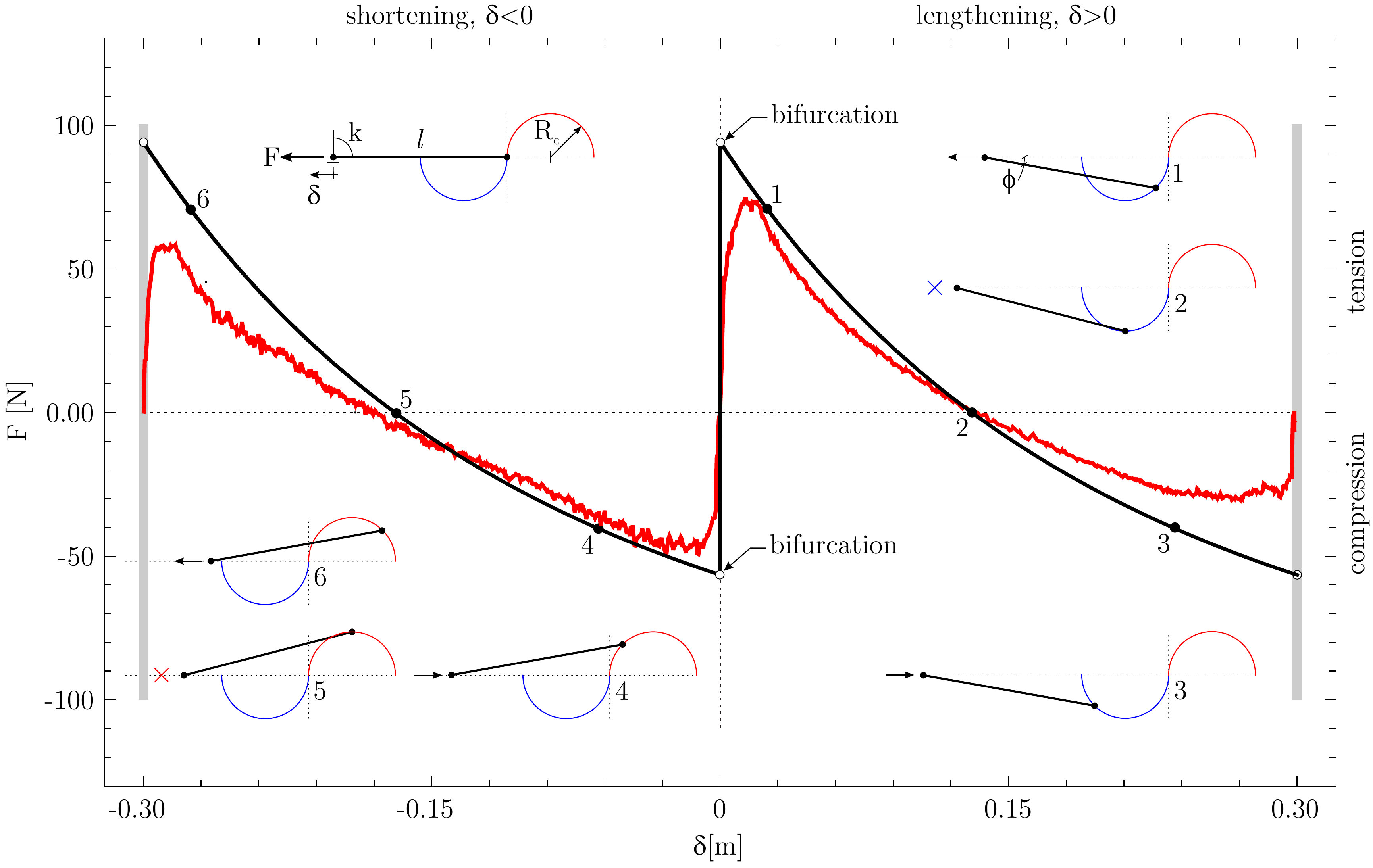}
    \caption{\footnotesize Load/dispacement experimental results (red line) versus theoretical prediction (black line) for a one-degree-of-freedom elastic structure having an elastic rotational hinge at the top and a roller constrained to slide on an \lq S-shaped', circular  profile
as shown in Fig.~\ref{ungrado}, together with the experimental set-up.
}
    \label{Experiments_DISCRET}
    \end{center}
\end{figure}

\begin{figure}[!htcb]
\renewcommand{\figurename}{\footnotesize{Fig.}}
    \begin{center}
    \includegraphics[width = 12 cm]{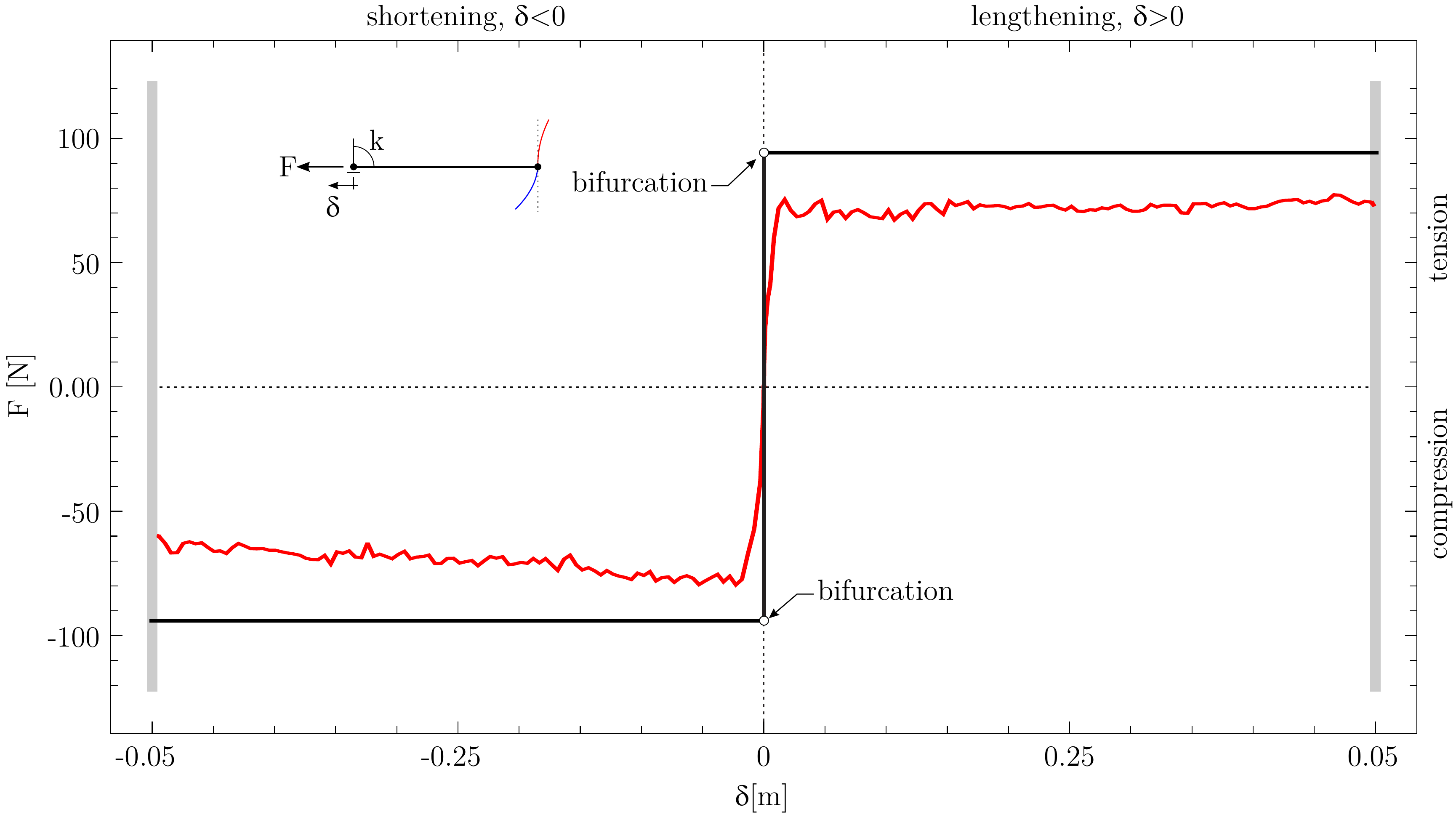}
    \caption{\footnotesize Load/dispacement experimental results (red line) versus theoretical prediction (black line) for a one-degree-of-freedom elastic structure designed to display a neutral postcritical behaviour.
The structure has an elastic rotational hinge at the top and a roller constrained to slide on the profile shown in Fig.~\ref{ungrado2}, together with the experimental set-up.
}
    \label{Experiments_POST_CONST}
    \end{center}
\end{figure}

Load/displacement curves are reported in Fig.~\ref{Experiments_DISCRET} for the \lq S-shaped' circular profile and in Fig.~\ref{Experiments_POST_CONST} for the profile giving the neutral response, as obtained from experiments, and directly compared with the theoretical predictions.
We note a nice agreement, with buckling detected prior to the attainment of the theoretical value, in
agreement with the known effect of the imperefections.
Friction at the roller/profile contact has induced some irrelevant load oscillation, minimized by hand-polishing the edges of the groove and using Areo Lubricant AS 100 (from Rivolta s.p.a, Milano, Italy).
We may finally comment that the experiments confirm the possibility of practically realizing mechanical systems behaving as the theoretical modelling predicts.

\section{The buckling and postcritical behaviour of an elastic structure with a circular constraint}\lb{CURV_EFF}

We consider an {\it inextensible} elastic rod (of bending stiffness $B$ and length $l$), with a movable clamp at one end, and having a rotational elastic spring (of stiffness $k$) on the other,
which can slide on a circle centered on the axis of the rod, see
the inset of Fig.~\ref{Graf_CRIT_LOADS}.
The rod is subject to an axial load $F$ which may be tensile ($F > 0$) or compressive ($F < 0$).

\subsection{The critical loads}\lb{VIB}

The linearized differential equilibrium equation of an elastic rod subject to an axial force $F$ is
\beq
\label{diff_equilibrium_eq}
\frac{d^4 v(z)}{dz^4} - \alpha^2\,\mbox{sgn}(F)\,\frac{d^2 v(z)}{dz^2} = 0 ,
\eeq
where $v$ is the transversal displacement, \lq sgn' is defined as sgn$(\alpha) = |\alpha|/\alpha$ $\forall \alpha \in \Re-\{0\}$, sgn$(0)=0$, and
\beq
\alpha^2 = \frac{|F|}{B}.
\eeq
The general solution of Eq.~(\ref{diff_equilibrium_eq}) is
\beq
\lb{sol_vib}
v(z) = \frac {C_1}{\alpha^2} \cosh(\sqrt{\mbox{sgn}(F)}\,\alpha \,z) + \frac {C_2}{\alpha^2}\sqrt{\mbox{sgn}(F)}\sinh(\sqrt{\mbox{sgn}(F)}\,\alpha \,z) + C_3 \,z + C_4 ,
\eeq
and the boundary conditions (involving the rotational spring stiffness $k$) are
\beq
\lb{bc_d_i}
v(0) = \frac{d v}{dz}\bigg|_{z=0} = 0 , ~~~
-\frac{\mbox{sgn}(F)}{\alpha^2} \frac{d^3 v}{dz^3}\bigg|_{z=l} = \phi + \ds \frac{d v}{dz}\bigg|_{z=l}, ~~~
-\frac{B}{k} \frac{d^2 v}{dz^2}\bigg|_{z=l} =  \phi + \ds \frac{d v}{dz}\bigg|_{z=l} ,
\eeq
plus the kinematic compatibility condition
\beq
\lb{bc_d_bend}
\phi=\widehat{\chi}/l v(l),
\eeq
involving the signed, dimensionless curvature $\widehat{\chi}=\pm l/R_c$ of the circle.

Imposing conditions (\ref{bc_d_i})--(\ref{bc_d_bend}), the solution (\ref{sol_vib}) provides the condition for the critical loads
\beqar
\barr{ll}
\lb{gen_Sol}
 \ds \big(\frac{1}{|\widehat{\chi}|}+\mbox{sgn}(\widehat{\chi})\big)\alpha\, l\, \mbox{sgn}(F)\cosh(\sqrt{\mbox{sgn}(F)}\,\alpha \,l)-\mbox{sgn}(\widehat{\chi})\sqrt{\mbox{sgn}(F)}\sinh(\sqrt{\mbox{sgn}(F)}\,\alpha \,l) {}  \nonumber\\ [5 mm]
+ \ds \frac{k}{ B\alpha } \left[\ds \big(\frac{1}{|\widehat{\chi}|}+\mbox{sgn}(\widehat{\chi})\big)\alpha\, l \sqrt{\mbox{sgn}(F)}\sinh(\sqrt{\mbox{sgn}(F)}\,\alpha \,l)+\mbox{sgn}(\widehat{\chi})\big(1-\cosh(\sqrt{\mbox{sgn}(F)}\,\alpha \,l)\big)\right]=0,
\earr
\eeqar
corresponding in the two limits $k \rightarrow 0$ and $k \rightarrow \infty$ to a pinned and clamped constraint on the right end, respectively.

\begin{figure}[!htcb]
\renewcommand{\figurename}{\footnotesize{Fig.}}
    \begin{center}
    \includegraphics[width = 12 cm]{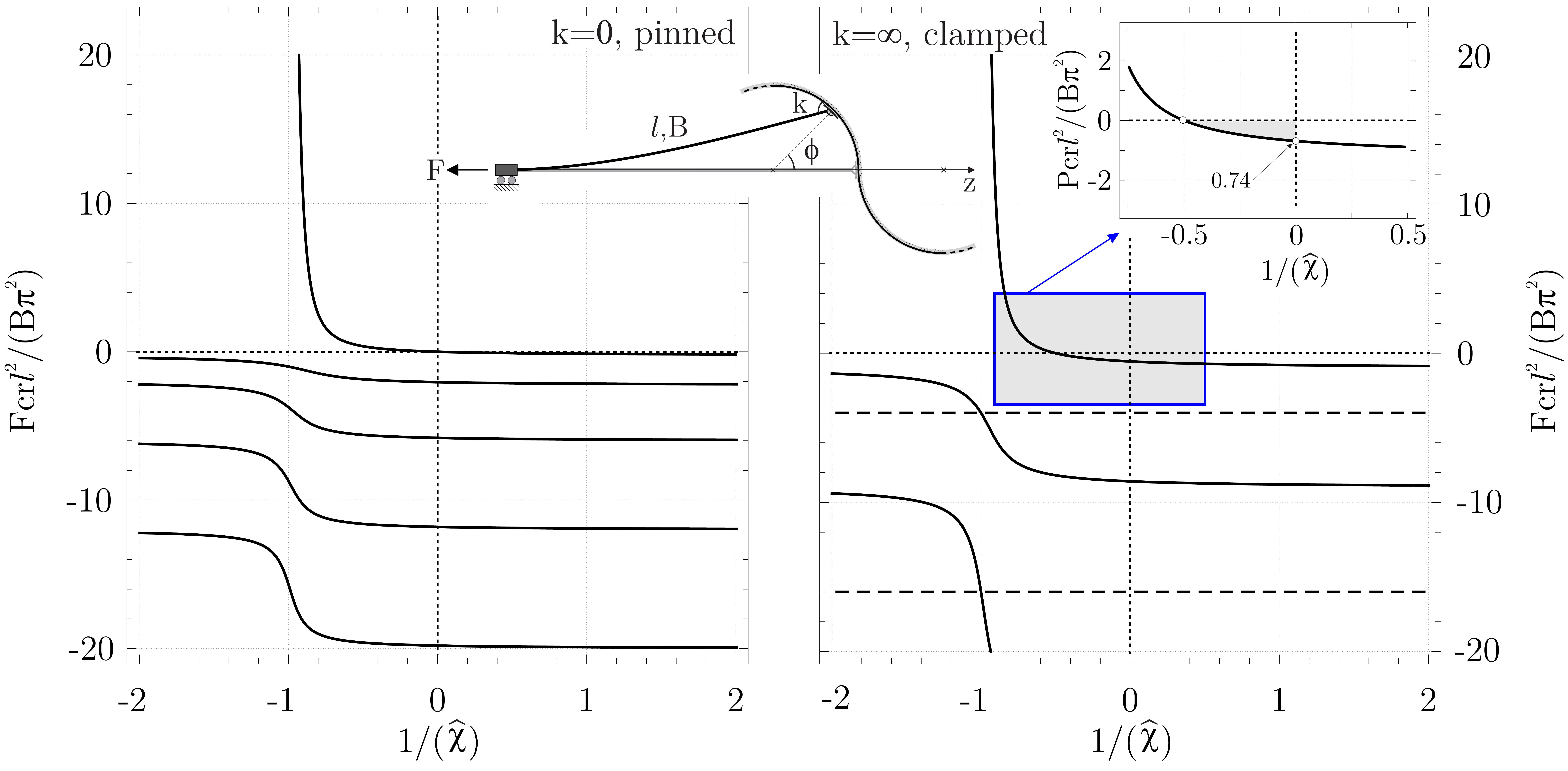}
    \caption{\footnotesize Dimensionless buckling load $F_{\rm{cr}}$ (a negative sign denotes compression) of the structure sketched in the inset (clamped on the left end and sliding along a circle on the right) as a
function of the signed dimensionless curvature $\widehat{\chi}$ of the circle.}
    \label{Graf_CRIT_LOADS}
    \end{center}
\end{figure}

\begin{figure}[!htcb]
\renewcommand{\tablename}{\footnotesize{Tab.}}
    \begin{center}
    \includegraphics[width = 12 cm]{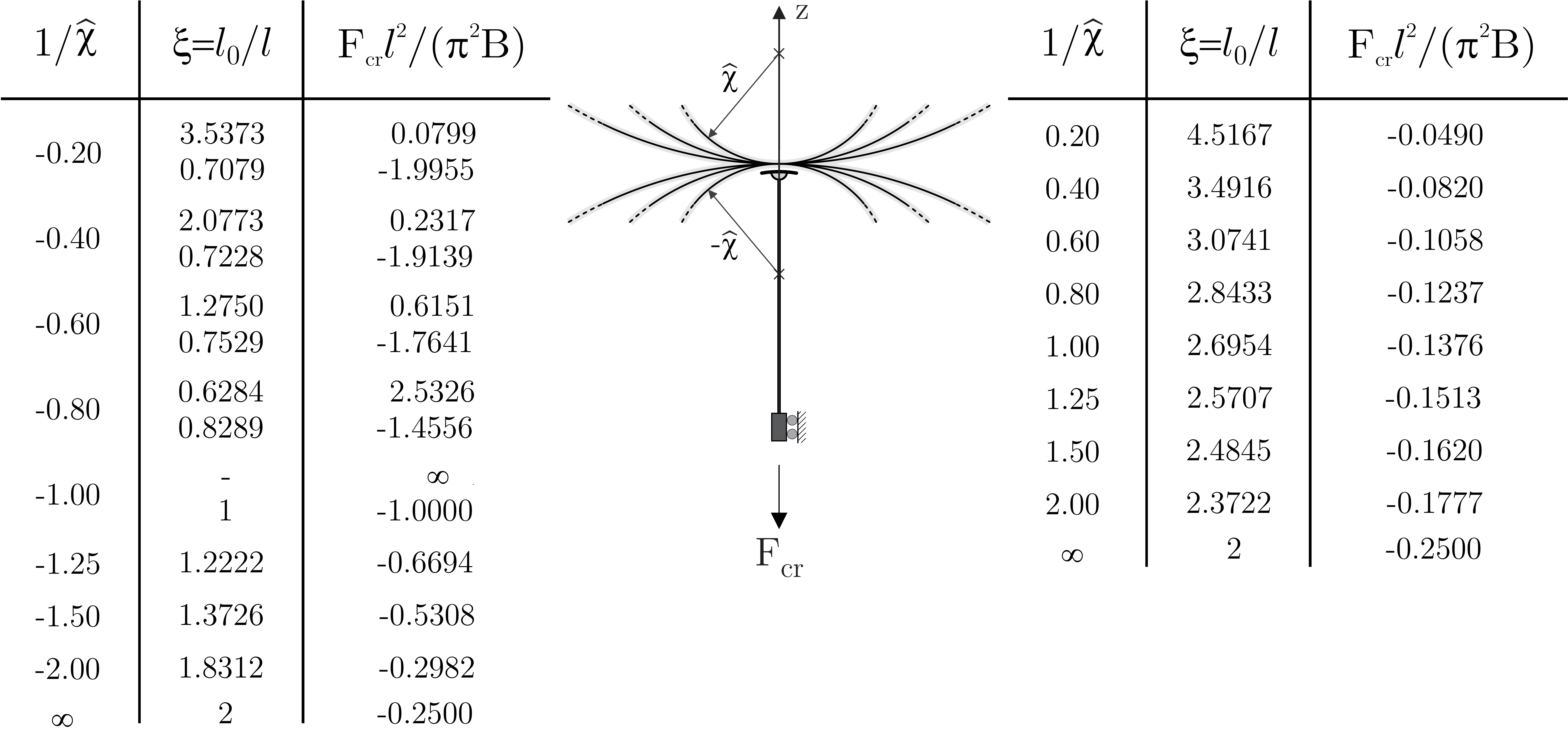}
    \ct{\footnotesize Dimensionless buckling load $F_{\rm{cr}}$ of the structure sketched in the inset (clamped on one end and sliding along a circle on the other) as a function of the signed dimensionless curvature $\widehat{\chi}$ of the circle. The structure is pinned on the right. A negative sign denotes a compressive load.}
    \label{Cr_loads_roller}
    \end{center}
\end{figure}

\begin{figure}[!htcb]
\renewcommand{\tablename}{\footnotesize{Tab.}}
    \begin{center}
    \includegraphics[width = 12 cm]{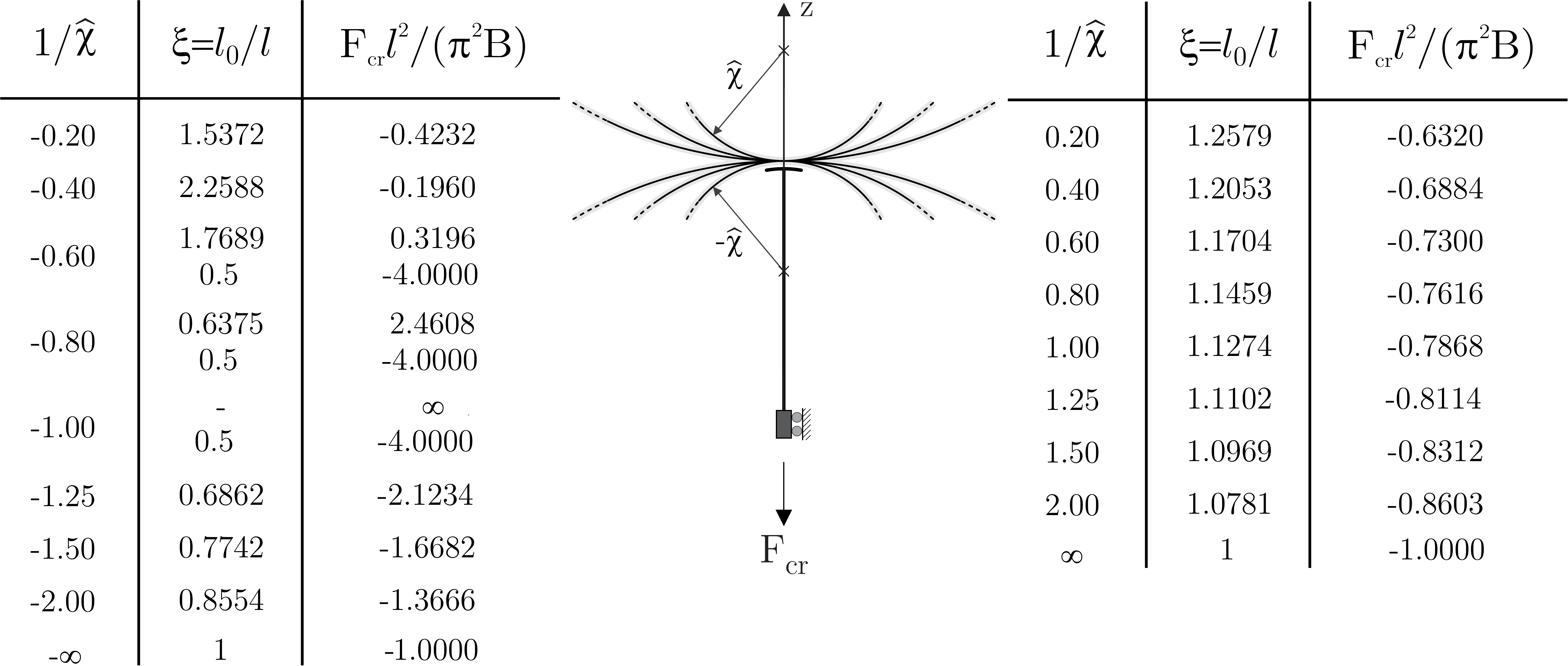}
    \ct{\footnotesize Dimensionless buckling load $F_{\rm{cr}}$ of the structure sketched in the inset (clamped on one end and sliding along a circle on the other) as a function of the signed dimensionless curvature $\widehat{\chi}$ of the circle. The structure is clamped on the right. A negative sign denotes a compressive load.}
    \label{Cr_loads_slider}
    \end{center}
\end{figure}

Buckling loads (made dimensionless through multiplication by $l^2/\pi^2/B$) are reported in Fig.~\ref{Graf_CRIT_LOADS} and in Tables \ref{Cr_loads_roller} and \ref{Cr_loads_slider}, as functions of the signed radius of curvature $\widehat{\chi}$ of the constraint.

Results reported in Tables \ref{Cr_loads_roller} and \ref{Cr_loads_slider} (where the negative signs denote compressive loads) are given in terms of effective length factor $\xi$ defined as
\beq
F_{cr} = \frac{\pi^2 B}{(\xi l)^2}.
\eeq

We note from the figure and from the tables that for certain curvatures of the constraint there is one buckling load in tension, while there are always infinite bifurcations
in compression (so that we can comment that the bifurcation problem remains a Sturm-Liouville problem).
The results reveal the strong effect of the constraint curvature, so that for instance for $ \widehat{\chi} =-1/0.2$ (for $ \widehat{\chi} =-1/0.8$)
there is a buckling load in tension much smaller (much higher) than that in compression (taken in absolute value), and for $\widehat{\chi} =-1/1.25$ (and for all positive curvatures $\widehat{\chi}>0$) there
is no tensile bifurcation.

\subsection{The elastica}\lb{ELASTICA}

The shape of the constraint also has a strong effect on the postcritical behaviour, as will be shown below with reference to the case of the circular profile.
Therefore, we derive the solution for an elastic rod clamped to the
left and constrained on the right to slide with a rotational spring (of
stiffness $k_r$) on a \lq S--shaped' bi-circular profile, as sketched in
Fig.~\ref{schema_elastica}, where the local reference system to be used in
the analysis is also indicated.
\begin{figure}[!htcb]
\renewcommand{\figurename}{\footnotesize{Fig.}}
    \begin{center}
    \includegraphics[width = 9 cm]{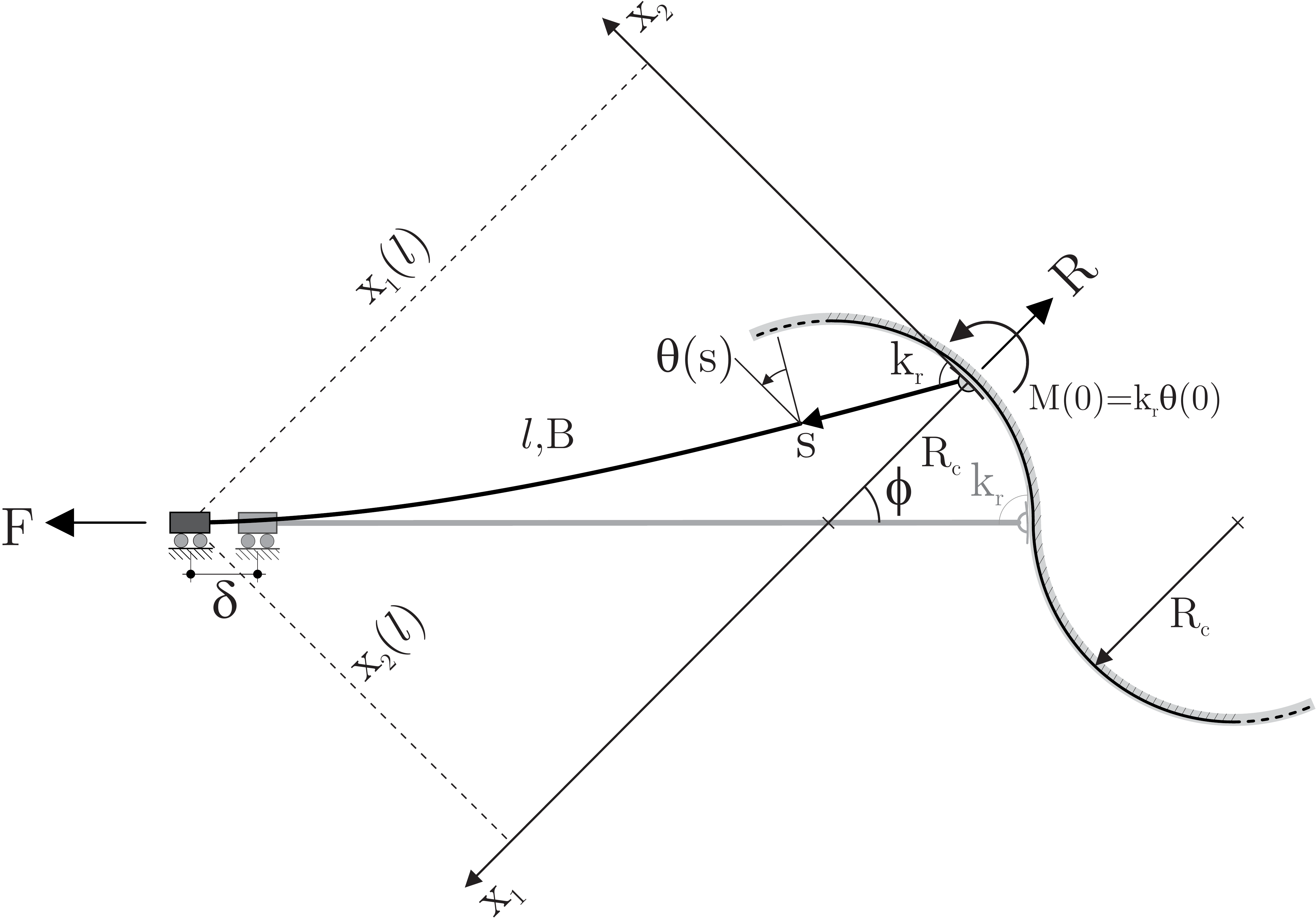}
    \caption{\footnotesize The elastic line problem for a rod clamped at the left end and constrained to slide with a hinge on a circle at the right end. Note the reference system employed in the analysis.}
    \lb{schema_elastica}
    \end{center}
\end{figure}

The elastic line problem is governed by the following equations.
\begin{enumerate}[i.)]

\item A condition of kinematic compatibility can be obtained by observing
from Fig.~\ref{schema_elastica} that the coordinates of the elastica
evaluated at $s = l$, namely, $x_1(l)$ and $x_2(l)$, are related to the
angle of rotation of the local reference system $\phi$ and to the radius
$R_c$ of the constraint via
    \beq
    \lb{cond_sol}
    \left[x_1(l) \mp R_c\right]\tan\phi - x_2(l) = 0,
    \eeq
    where $\phi$ is assumed positive if anticlockwise; note that in
Eq.~(\ref{cond_sol}) the sign \lq $-$' (\lq $+$') holds for the case of pin
lying on the left (right) half--circle.

\item The curved constraint transmits to the rod a moment and a force
pointing the center of the circle, in other words, parallel to $x_1$ and
assumed positive when opposite to the direction of the $x_1$--axis, so that
for $0 \le \phi < \pi/2$ ($\pi/2 < \phi \le \pi$) it corresponds to a
positive tensile (negative compressive) dead force $F$ applied to the
structure defined by
    \beq
    \lb{F_R}
    F = R\cos\phi.
    \eeq

\item Through introduction of the curvilinear coordinate $s$, the elastica governing
deflections of the rod is
    \beq
    \lb{eq_ind}
    \frac{d^2\theta}{ds^2} - \frac{R}{B} \sin\theta = 0,
    \eeq
    where $\theta$ is the rotation angle (assumed positive if clockwise) of
the normal at each point of the elastica, so that with the symbols
introduced in Fig.~\ref{schema_elastica} it is
    \beq
    \lb{bc_l}
    \theta(l) = \phi.
    \eeq

\end{enumerate}

Integration of Eq.~(\ref{eq_ind}) from $0$ to $s$, after multiplication by
$d\theta / ds$, leads to \beq \lb{first_int}
\left(\frac{d\theta}{ds}\right)^2 = 2\,\tilde{\alpha}^2\left[\frac{2}{k^2}
- 1 - \mbox{sgn}(R)\cos\theta\right] , \eeq where \beq \lb{k_value}
\tilde{\alpha}^2 = \frac{|R|}{B} ,~~~~~~~ k^2 = \ds
\frac{4\tilde{\alpha}^2}{\left[\theta(0)\,k_r/B\right]^2 +
2\tilde{\alpha}^2\left[\mbox{sgn}(R)\cos\theta(0) + 1\right]} , \eeq in
which the term $\theta(0)\,k_r/B$ corresponds to the curvature of the rod
evaluated at $s = 0$. The introduction of the change of variable \beq \beta
= [\theta - \mbox{H}(R)\,\pi]/2, \eeq where $\mbox{H}$ denotes the
Heaviside step function, allows to re-write Eq.~(\ref{first_int}) as \beq
\left(\frac{d\beta}{ds}\right)^2 = \ds \frac{\tilde{\alpha}^2}{k^2} \left(1
- k^2\sin^2\beta\right) , \eeq so that a second change of variable $u = s
\tilde{\alpha} / k$ yields \beq \lb{dbe_du} \frac{d\beta}{du} = \ds
\pm\sqrt{1 - k^2\sin^2\beta} . \eeq

Restricting the treatment to the case \lq$+$', which corresponds to
$\theta(0)\geq0$, and since $\beta = \beta(0)$ at $u=0$, Eq.~(\ref{dbe_du})
provides the following solution for $\beta$
\beq
\lb{be_sol}
\beta = \ds
\mbox{am}\left[u + \mbox{F}\left[\beta(0), k\right], k\right],
\eeq
where
am and F are the Jacobi elliptic function amplitude and the incomplete
elliptic integral of the first kind of modulus $k$, respectively (Byrd and
Friedman, 1971). Keeping into account that $dx_1/ds = \cos\theta$ and
$dx_2/ds = \sin\theta$, an integration provides the two coordinates $x_1$
and $x_2$ of the elastica expressed in terms of $u$ as
\beq \lb{cor} \left\{
\barr{l} \ds x_1 = \mbox{sgn}(R)\,\frac{2}{k\tilde{\alpha}} \{(1 - k^2/2)u
+ \mbox{E}\left[\beta(0), k\right] - \mbox{E}\left[\mbox{am}\left[u +
\mbox{F}\left[\beta(0), k\right], k\right], k\right]\} , \\[6mm] \ds x_2 =
\mbox{sgn}(R)\,\frac{2}{k\tilde{\alpha}} \{\mbox{dn}\left[u +
\mbox{F}\left[\beta(0), k\right], k\right] -
\mbox{dn}\left[\mbox{F}\left[\beta(0), k\right], k\right]\} , \earr \right.
\eeq in which the constants of integration are chosen so that $x_1$ and
$x_2$ vanish at $s = 0$. In Eqs.~(\ref{cor}) dn is the Jacobi elliptic
function delta-amplitude of modulus $k$, while E is the incomplete elliptic
integral of the second kind (Byrd and Friedman, 1971). Since $\theta(0)$
may be now not null, Eqs.~(\ref{cor}) generalize the expressions derived by
Zaccaria {\it et al.} [2011, their equations (3.23) and (3.24)].

The horizontal displacement $\delta$ of the clamp on the left of the
structure (assumed positive for a lengthening of the system) is given by
the form \beq \lb{h_disp} \delta = \frac{x_2}{\sin\phi} - l \mp R_c, \eeq
where as for Eq.~(\ref{cond_sol}) the sign \lq $-$' (\lq $+$') holds for
the case of pin lying on the left (right) half--circle.

The axial load $F$ can be obtained as a function of the rotation $\phi$, or
as a function of the end displacement $\delta$, by following the steps
below: \begin{enumerate}[i.)]

\item a value for $\theta(0)$ is fixed, so that $k$ can be expressed using
Eqs.~(\ref{k_value}) as a function of $R$;

\item the expressions~(\ref{cor}) for the coordinates of the elastica and
Eq.~(\ref{be_sol}), evaluated at $s = l$, become functions of $R$ only;

\item Eq.~(\ref{bc_l}) provides $\phi$, so that Eq.~(\ref{cond_sol})
becomes a nonlinear equation in the variable $R$, which can be numerically
solved (we have used the function FindRoot of
$\mbox{Mathematica}^{\mbox{\tiny\textregistered}}$~6.0);

\item once $R$ is known, $F$, $\phi$ and $\delta$ can be respectively
obtained from Eqs.~(\ref{F_R}), (\ref{bc_l}) and (\ref{h_disp}).

\end{enumerate}

The postcritical behaviour (branching from both tensile and compressive
critical loads) of the structure is reported in Fig.~\ref{bif_tra} in terms
of dimensionless axial load $4Fl^2/(B\pi^2)$ versus dimensionless
displacement $\delta/R_c$ for the particular case of a roller sliding on the profile, $k_r = 0$.
\begin{figure}[!htcb]
\renewcommand{\figurename}{\footnotesize{Fig.}}
    \begin{center}
    \includegraphics[width = 12 cm]{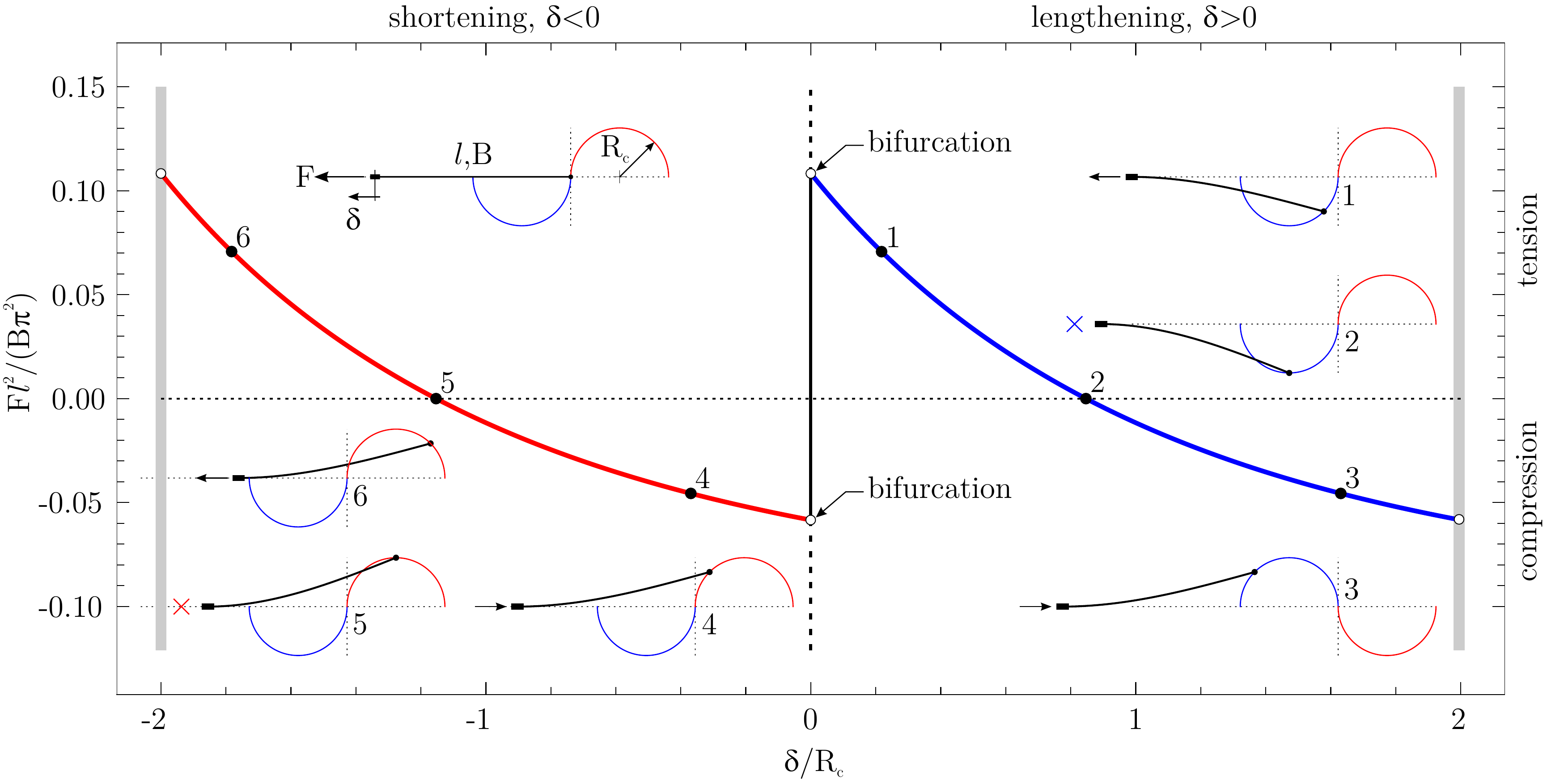}
    \caption{\footnotesize
The postcritical behaviour of the structure sketched in the inset (with a roller sliding on the \lq S-shaped' profile) under tensile and compressive loads. Dimensionless axial load $F$ versus dimensionless end displacement.}
    \lb{bif_tra}
    \end{center}
\end{figure}

We note that the elastica obtained in this case is {\it inflexional} and therefore different from that reported by Zaccaria {\it et al.} (2011),
moreover, the postcritical behaviour is always unstable,
evidencing decrease of the load with increasing edge displacement (\lq softening').
Special features of the postcritical behaviour (already present in the one-degree-of-freedom system) are (i.) that there is a transition from a tensile (compressive) to compressive (tensile) elastica when the
constraint reaches the points denoted with \lq 2' and \lq 5' in the graph, and that (ii.) the postcritical branches emanating from the critical loads are the same, but horizontally shifted.

\subsection{Experiments on the elastica}\lb{EXPERIMENT}

\begin{figure}[!htcb]
\renewcommand{\figurename}{\footnotesize{Fig.}}
    \begin{center}
    \includegraphics[width = 12 cm]{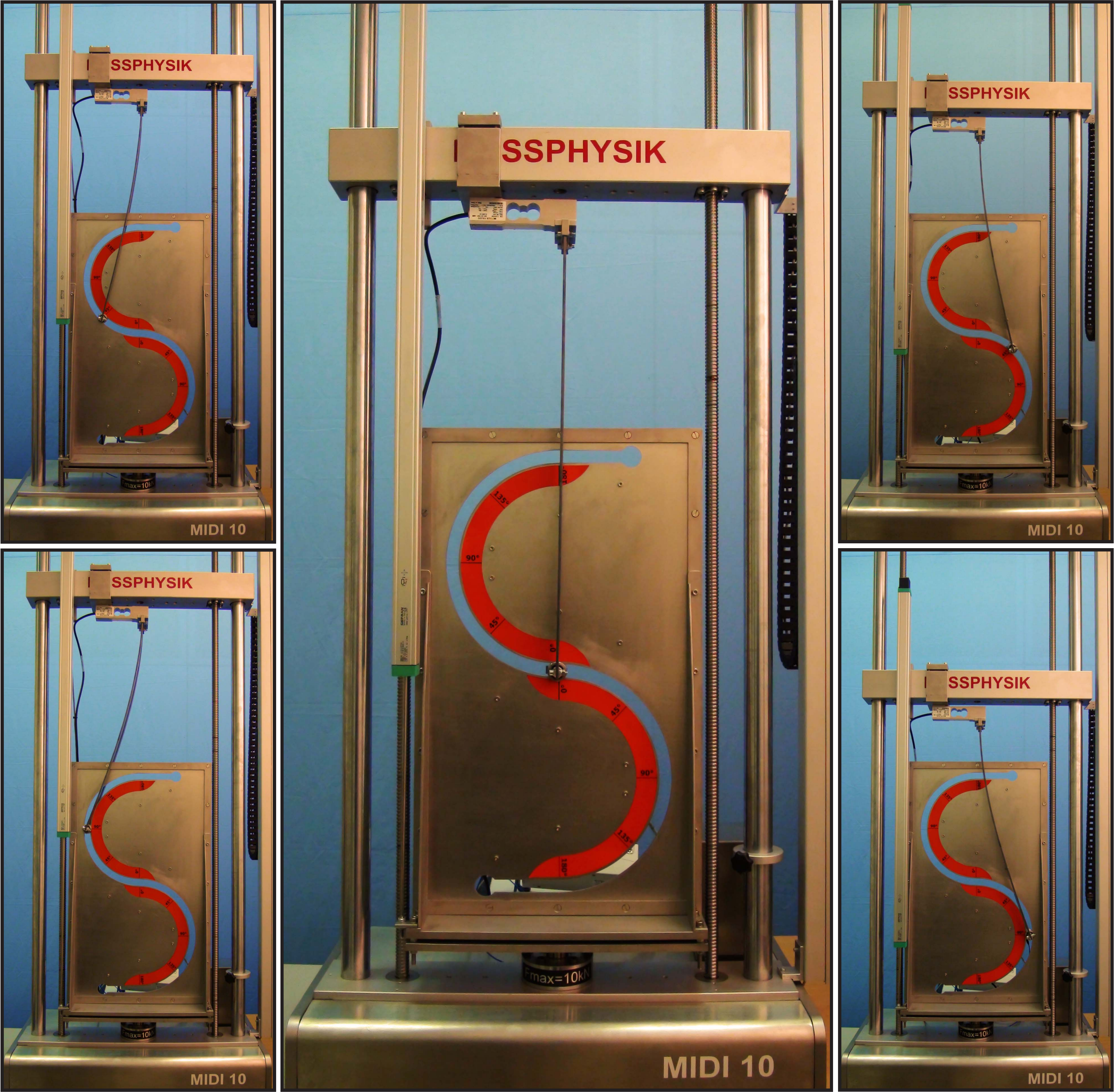}
    \caption{\footnotesize Deformed shapes of the elastica during a test of a beam sliding on an \lq S-shaped', circular  profile. Two photos taken during elongation (shortening) are reported on the left (on the right).}
    \label{confronto-deformate}
    \end{center}
\end{figure}

We have tested the behaviour of a continuos system by employing the same experimental set-up used for testing the one-degrees-of-freedom structures in Sect. \ref{exp-123}, but with the rigid system replaced by elastic rods
realized with two 250 mm $\times$ 25 mm $\times$ 4 mm C72 carbon-steel strips (Young modulus 200 GPa, weight 968\,gr), see Appendix A for details. The experimental set-up with photos taken during the tests is shown in Fig. \ref{confronto-deformate}.

\begin{figure}[!htcb]
\renewcommand{\figurename}{\footnotesize{Fig.}}
    \begin{center}
    \includegraphics[width = 12 cm]{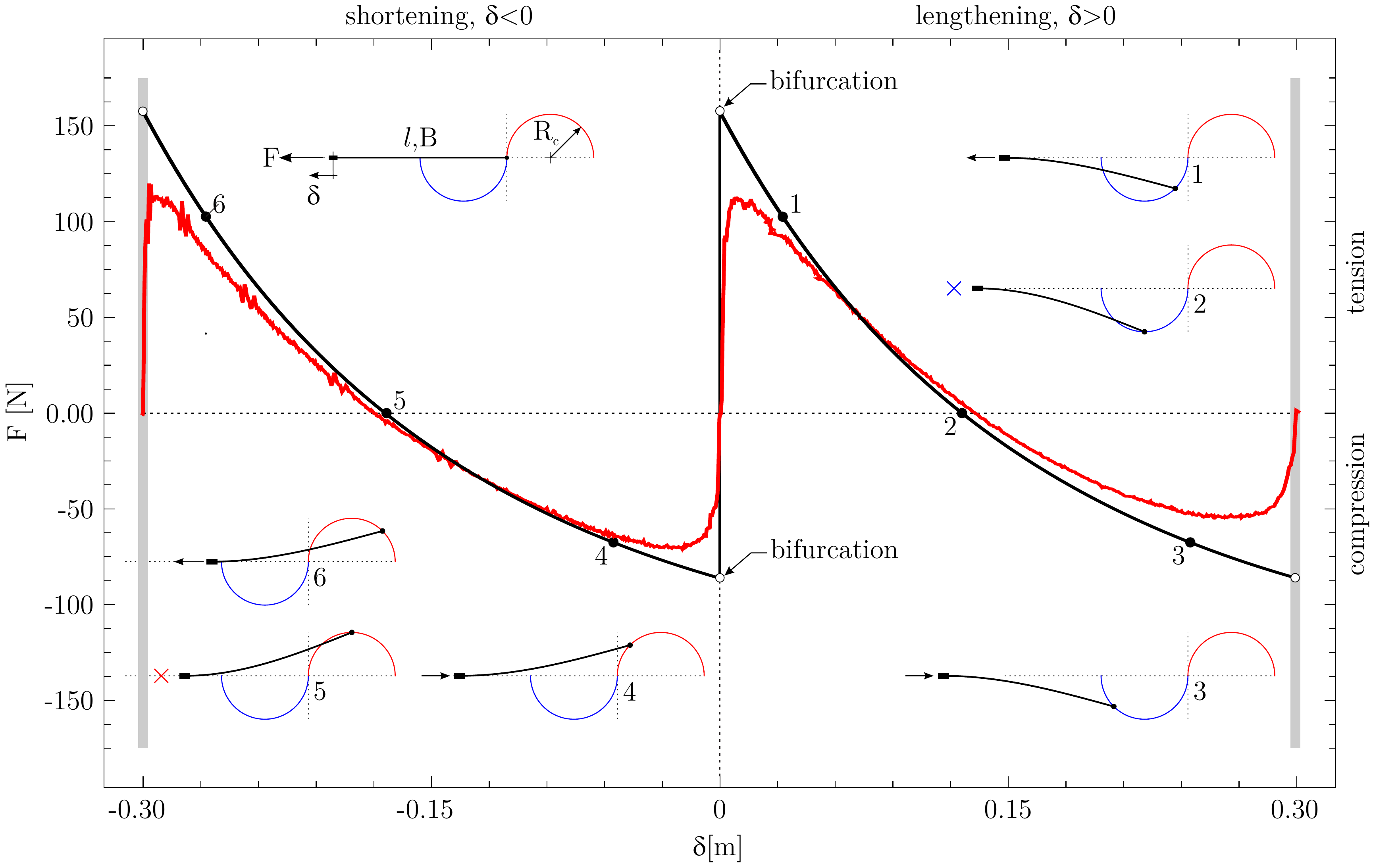}
    \caption{\footnotesize
Load/dispacement experimental results (red line) versus theoretical prediction (black line) for
a beam sliding on an \lq S-shaped', circular profile (see the inset).
}
    \label{Experiments_CONT}
    \end{center}
\end{figure}

\begin{figure}[!htcb]
\renewcommand{\figurename}{\footnotesize{Fig.}}
    \begin{center}
    \includegraphics[width = 14 cm]{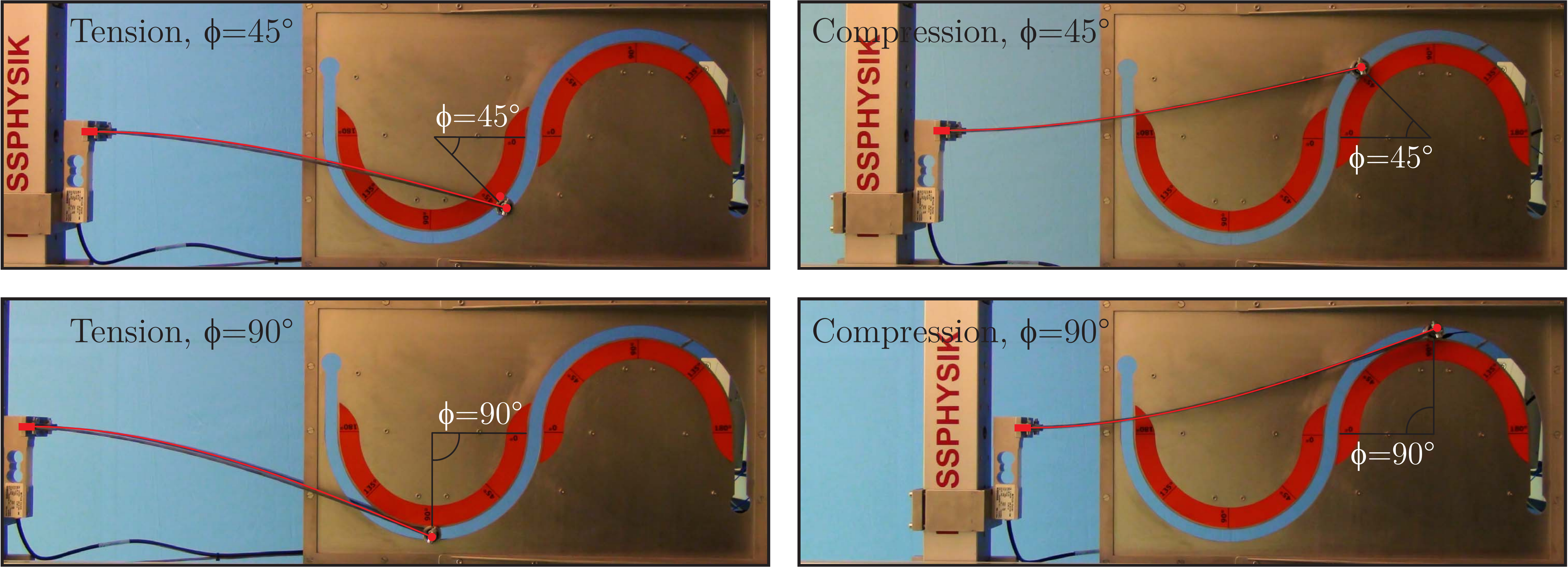}
    \caption{\footnotesize Deforemd shapes of the elastica during a test of a beam sliding on an \lq S-shaped', circular profile. This figure provides a direct comparison between
details taken from Fig.~\ref{confronto-deformate} and the predictions of the elastica (reported with a red line).}
    \label{confrontissimo}
    \end{center}
\end{figure}

Experimental results are reported in Fig. \ref{Experiments_CONT}
in terms theoretical (red line) versus experimental (black line) force/end displacement data.
Moreover, the photos reported in Fig.~\ref{confrontissimo}, which are details of the photos shown in Fig.~\ref{confronto-deformate} on the left and on the right,
are compared with the
theoretical elastica [shown red and obtained from Eqs.~(\ref{cor}))] at four different end angles (45$^\circ$ and 90$^\circ$ for tension and compression).

From the figures, we can observe the following facts.

\begin{itemize}
\item The experiments definitely substantiate theoretical findings.

\item The comparison between the deformed beam during a test and the predictions of the elastica, shown in Fig.~\ref{confrontissimo}, reveals a very tight agreement between theory and experiments.

\end{itemize}

As for the one-degree-of-freedom systems, we can again conclude that the experiments confirm the possibility of practically realizing elastic systems behaving in strict agreement with theoretical predictions.

\section*{Conclusions}

Effects related to the curvature and the shape of the constraint profile on which an end of a structure has to slide have been shown to be important on bifurcation and instability.
In particular, we have found possibility of buckling in tension and compression and also multiple buckling loads, as for instance in the case of a one-degree-of-freedom structure evidencing two critical loads.
Our experiments have confirmed that these effects can be designed to occur in real structural prototypes, so that
new possibilities are opened in exploiting simple deformational mechanisms to obtain flexible mechanical systems.

\section*{Acknowledgments}
D.B. and G.N. gratefully acknowledge financial support from Italian Prin 2009 (prot. 2009XWLFKW-002); D.B. also acknowledges support from grant PIAP-GA-2011-286110.


\appendix
\renewcommand\thesection{Appendix \Alph{section}}
\renewcommand\thesubsection{\Alph{section}.\arabic{subsection}}

\numberwithin{equation}{section}
\renewcommand{\theequation}{\Alph{section}.\arabic{equation}}

\section{Details on the experimental setup}\lb{APP_A}

Experiments reported in the present article have been performed at the Laboratory for Physical Modeling of Structures and Photoelasticity of the University of Trento (managed by D.B.).
A Midi 10 (10 KN maximum force, from Messphysik Materials Testing) electromechanical testing machine has been employed to impose displacements (velocity 0.2 mm/s) at
the ends of the structures.
Loads and displacements have been measured with the loading cell and the displacement transducers mounted on the Midi 10 machine, and, independently, with a MT 1041
(0.5 kN maximum load) load cell
(from Mettler-Toledo)
and displacement with a potentiometric
transducer Gefran LTM-900-S IP65 (from Gefran Spa).

The rotational springs employed for the one-degree-of-freedom systems have been designed
to provide a stiffness equal to 211.5 Nm by employing equations (32) of Brown (1981). After machining, the springs have been tested and found to correspond to a stiffness equal to 169.5 Nm, the value which has been used to compare experiments
with theoretical results.

An IEPE accelerometer (PCB Piezotronics Inc., model 333B50) has been attached at one end of the structure to detect the instant of buckling.
This has been observed in all the tests to correspond to an acceleration peak ranging between 0.15 and 0.2 g, while before buckling and during postcritical behaviour the acceleration
did not exceede the value 0.003\,g.

Data from the MT 1041, the Gefran LTM-900-S IP65 and the accelerometer
have been acquired with system NI CompactDAQ,
interfaced with Labview 8.5.1 (National Instruments), while acquisition of the data from the Midi 10 has been obtained from Doli controller (from Messphysik).

Temperature near the testing machine has been monitored with a thermocouple
connected to a Xplorer GLX (from Pasco) and has been found to lie
around 22$^\circ$C, without sensible oscillations during tests.

\begin{figure}[H]
\renewcommand{\figurename}{\footnotesize{Fig.}}
    \begin{center}
    \includegraphics[width = 6 cm]{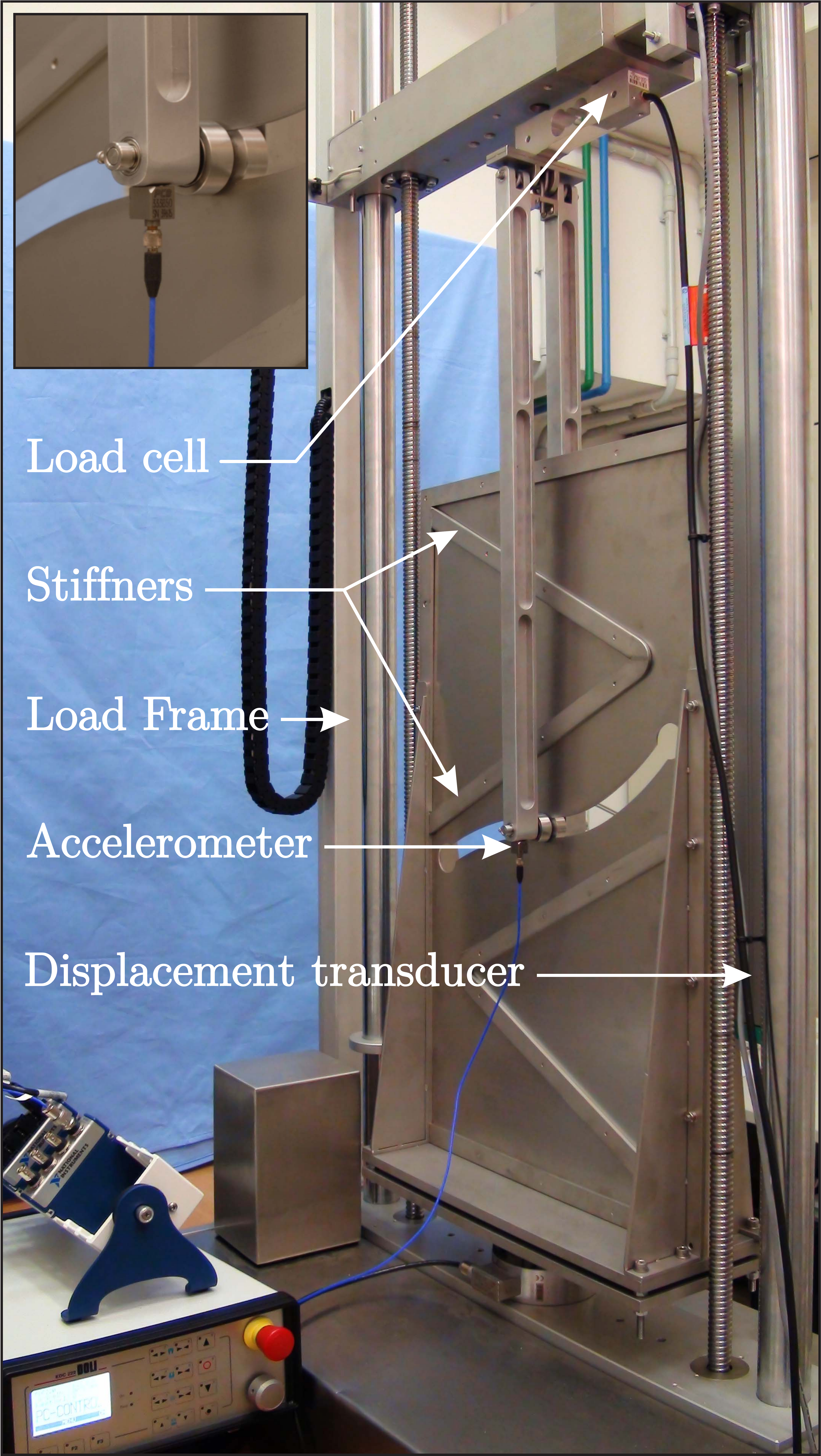}
    \caption{\footnotesize The experimental setup for the buckling tests, seen from the back.}
    \label{test_setup}
    \end{center}
\end{figure}

Photos have been taken with a Nikon D200 digital camera, equipped with AF Nikkor (18-35mm 1:3.5-4.5 D) lens (Nikon Corporation) and
movies have been recorded during the tests with a Sony handycam (model HDR-XR550VE). The testing setup is shown in Fig. \ref{test_setup}. Additional material can be found at
http://www.ing.unitn.it/dims/ssmg.php.


\begin{thebibliography}{99}

\setlength{\itemsep}{-1.0mm}

{\footnotesize

\bibitem{spring} Brown A.A.D. (1981) {\it Mechanical springs}. Oxford University Press.

\bibitem{byrd} Byrd, P.F and Friedman, M.D. (1971) {\it Handbook of elliptic integrals for engineers and scientists}. Springer-Verlag.

\bibitem{gaspar}  G\'{a}sp\'{a}r, Zs. (1984) Buckling model for a degenerated case. {\it News Letter of the Technical University of Budapest},  4, pp. 5–8.

\bibitem{love} Love, A.E.H. (1927) {\it A treatise on the mathematical theory of elasicity}. Cambridge University Press.

\bibitem{tim} Timoshenko S.P. and Gere, J.M. (1936) {\it Theory of elastic stability}. McGraw-Hill.

\bibitem{zak} Zaccaria, D., Bigoni, D., Noselli, G. and Misseroni, D. (2011) Structures buckling under tensile dead load. \PRSL A, 2011, 467, 1686-1700.

}

\end{thebibliography}
\end{document}